# Global-scale massive feature extraction from monthly hydroclimatic time series: Statistical characterizations, spatial patterns and hydrological similarity

Georgia Papacharalampous[1,2,3,*], Hristos Tyralis[3,4], Simon Michael Papalexiou[5,6,7], Andreas Langousis[2], Sina Khatami[8,9,10], Elena Volpi[1], and Salvatore Grimaldi[11,12]

1. Department of Engineering, Roma Tre University, Rome, Italy
2. Department of Civil Engineering, School of Engineering, University of Patras, University Campus, Rio, 26504, Patras, Greece
3. Department of Water Resources and Environmental Engineering, School of Civil Engineering, National Technical University of Athens, Heroon Polytechneiou 5, 15780 Zographou, Greece
4. Air Force Support Command, Hellenic Air Force, Elefsina Air Base, 19200 Elefsina, Greece
5. Department of Civil, Geological and Environmental Engineering, University of Saskatchewan, Saskatoon, Saskatchewan, Canada
6. Global Institute for Water Security, Saskatoon, Saskatchewan, Canada
7. Faculty of Environmental Sciences, Czech University of Life Sciences Prague
8. Department of Physical Geography and the Bolin Centre for Climate Research, Stockholm University, SE-10691 Stockholm, Sweden
9. Climate & Energy College, University of Melbourne, Parkville, Victoria, Australia
10. Department of Infrastructure Engineering, University of Melbourne, Parkville, Victoria, Australia
11. Department for Innovation in Biological, Agro-food and Forest Systems, University of Tuscia, Viterbo, Italy
12. Department of Mechanical and Aerospace Engineering, Tandon School of Engineering, New York University, Brooklyn, NY, 10003, USA

* Corresponding author (papacharalampous.georgia@gmail.com)

**Email addresses and ORCID profiles:** papacharalampous.georgia@gmail.com, geopap@upatras.gr, gpapacharalampous@hydro.ntua.gr, https://orcid.org/0000-0001-5446-954X (Georgia Papacharalampous); montchrister@gmail.com, hristos@itia.ntua.gr, https://orcid.org/0000-

0002-8932-4997 (Hristos Tyralis); sm.papalexiou@usask.ca, https://orcid.org/0000-0001-5633-0154 (Simon Michael Papalexiou); andlag@alum.mit.edu, https://orcid.org/0000-0002-0643-2520 (Andreas Langousis); sina.khatami@natgeo.su.se, https://orcid.org/0000-0003-1149-5080 (Sina Khatami); elena.volpi@uniroma3.it, https://orcid.org/0000-0002-9511-1496 (Elena Volpi); salvatore.grimaldi@unitus.it, http://orcid.org/0000-0001-5715-106X (Salvatore Grimaldi)

**Abstract:** Hydroclimatic time series analysis focuses on a few feature types (e.g., autocorrelations, trends, extremes), which describe a small portion of the entire information content of the observations. Aiming to exploit a larger part of the available information and, thus, to deliver more reliable results (e.g., in hydroclimatic time series clustering contexts), here we approach hydroclimatic time series analysis differently, i.e., by performing massive feature extraction. In this respect, we develop a big data framework for hydroclimatic variable behaviour characterization. This framework relies on approximately 60 diverse features and is completely automatic (in the sense that it does not depend on the hydroclimatic process at hand). We apply the new framework to characterize mean monthly temperature, total monthly precipitation and mean monthly river flow. The applications are conducted at the global scale by exploiting 40-year-long time series originating from over 13 000 stations. We extract interpretable knowledge on seasonality, trends, autocorrelation, long-range dependence and entropy, and on feature types that are met less frequently. We further compare the examined hydroclimatic variable types in terms of this knowledge and, identify patterns related to the spatial variability of the features. For this latter purpose, we also propose and exploit a hydroclimatic time series clustering methodology. This new methodology is based on Breiman's random forests. The descriptive and exploratory insights gained by the global-scale applications prove the usefulness of the adopted feature compilation in hydroclimatic contexts. Moreover, the spatially coherent patterns characterizing the clusters delivered by the new methodology build confidence in its future exploitation. Given this spatial coherence and the scale-independent nature of the delivered feature values (which makes them particularly useful in forecasting and simulation contexts), we believe that this methodology could also be beneficial within regionalization frameworks, in which knowledge on hydrological similarity is exploited in technical and operative terms.

# 1. Introduction

## 1.1 Representations and characterizations of hydroclimatic variables

The study of hydroclimatic variables and related topics and themes —such as precipitation, temperature and river flow dynamics— across spatio-temporal scales is a strategic research priority in a changing world. In line with this priority, pathways of hydroclimatic research are largely focused on the underlying mechanisms of hydroclimatic variables, their role in the sustenance of Earth systems, their changes, inter-relationships, and relationships with climatic regimes and catchment characteristics (see, e.g., Montanari et al., 2013; Blöschl et al., 2019a; Fan et al., 2019). In this context, representations and characterizations can be exploited to advance our understanding and scientific models of this important family of real-world variables.

Representations and characterizations can take various formulations not only as the core of diagnostic and exploratory frameworks, but also as the basis for prediction methodologies of all possible types (see, e.g., Pechlivanidis et al., 2014; Tyralis and Koutsoyiannis, 2017; Tyralis et al., 2020), and as the basis for hydrological and hydroclimatic (stochastic) simulation frameworks (see, e.g., Perrin et al., 2003; Kumar et al., 2006; Langousis and Koutsoyiannis, 2006; Lee and Salas, 2011; Grimaldi et al., 2012; Langousis and Kaleris, 2014; Papalexiou, 2018). The candidate formulations may include (but are not limited to) statistical characterizations and representations in terms of marginal probability (or cumulative) density functions (see, e.g., Kroll et al., 2002; Papalexiou and Koutsoyiannis, 2013; Nerantzaki and Papalexiou, 2019), joint probability density functions and copulas (see, e.g., Serinaldi et al., 2009; Kuchment and Demidov, 2013; Wong et al., 2013), time series or regression models (see, e.g., Carlson et al., 1970; Koutsoyiannis, 2011; Khatami, 2013; Khazaei et al., 2019; Papalexiou and Montanari, 2019; Ghajarnia et al. 2020; Kagawa-Viviani and Giambelluca, 2020), process-based (including conceptual) representations (see, e.g., the reviews in Langousis and Koutsoyiannis, 2006; Koutsoyiannis and Langousis, 2011; Jaramillo and Destouni 2015; Langousis et al., 2016a; Davtalab et al., 2017; Széles et al., 2018; Khatami et al., 2019; Emmanouil et al., 2020; Khatami et al., 2020), and characterizations through feature extraction; see Section 1.2 below.

## 1.2 Features of hydroclimatic variables: Definition, extensions and examples

In their typical form, features (also known as "signatures", "statistical characteristics", or simply "statistics") are sample statistics or model parameter estimates that can summarize or measure specific properties of real-world processes. They can also be exploited in a straightforward way (i.e., through regression analyses) for identifying important relationships between such properties. For instance, the Hurst parameter of the fractional Gaussian noise (fGn) process —initially introduced by Kolmogorov (1940), extensively studied by Mandelbrot, Wallis, and Van Ness in the 1960s (see, e.g., Mandelbrot and Wallis, 1968), and popularized in recent works (e.g., Beran, 1994; Koutsoyiannis, 2002; Montanari, 2003; O'Connell et al., 2016)— is a feature indicating the magnitude of long-range dependence (when computed for non-seasonal processes). Global-scale investigations on this dependence and its relationships with other statistical properties of annual precipitation and annual runoff can be found in Tyralis et al. (2018; see also the references therein) and Markonis et al (2018b), respectively. Similarly, the shape parameter of the generalized extreme value distribution, when the latter is fitted to daily annual block maxima of streamflow, can serve as an indicator for flood extremity (see, e.g., the investigations for North America by Tyralis et al. 2019c and the references therein).

By extension, less interpretable descriptive features (i.e., any statistic or model parameter estimate, independently of its interpretability) may also be relevant for various purposes and tasks (see, e.g., Fulcher and Jones, 2014; Christ et al., 2018; Fulcher, 2018; Lubba et al., 2019), including feature-based time series clustering (distinguished from other time series clustering approaches in, e.g., Aghabozorgi et al., 2015). Moreover, process predictability features can be extracted by characterizing the out-of-sample predictive performance (of statistical or process-based models) in terms of selected scores (see, e.g., the global-scale predictability characterizations of monthly temperature and precipitation in terms of the Nash-Sutcliffe efficiency by Papacharalampous et al., 2018). In this view, feature extraction facilitates the study of hydroclimatic (and other geophysical) processes from both the descriptive and predictive perspectives. These two perspectives, distinguished on a theoretical basis in Shmueli (2010), have important technical and operative implications, and may be linked (to a larger or smaller extent) to each other (see, e.g., the investigations for North America and Europe on the relationships between selected predictability and descriptive annual river flow features in Papacharalampous and Tyralis, 2020, and the detailed study on the key hydroclimatic and physiographic drivers controlling seasonal river flow predictability across Europe by Pechlivanidis et al., 2020). In what follows, and unless specified differently, we will refer to descriptive features of hydroclimatic variables, obtained through statistical analyses in geoscience, since the present work is devoted to such features.

### 1.3 Features of hydroclimatic variables: Brief overview and investigated concepts

Features are regularly computed and studied in many diverse branches of geoscience, including statistical-stochastic geoscience (see, e.g., Montanari et al., 1997; Grimaldi, 2004; Koutsoyiannis, 2011; Volpi and Fiori, 2012; Markonis and Koutsoyiannis, 2013; Papalexiou et al., 2013; Volpi et al., 2015; Villarini, 2016; Markonis et al., 2018b; Tyralis et al., 2018; Volpi, 2019; Marra et al., 2020; Serinaldi et al., 2020a) and catchment hydrology (see, e.g., Pallard et al., 2009; Baratti et al., 2012; Euser et al., 2013; Toth, 2013; Zhang et al., 2014; Shafii and Tolson, 2015; Westerberg and McMillan, 2015; Boscarello et al., 2016; Donnelly et al., 2016; Su et al., 2016; Westerberg et al., 2016; Fang et al., 2017; McMillan et al., 2017; Addor et al., 2018; Zhang et al., 2018). This latter field (in which features are referred to as "hydrological signatures") shows the wide applicability of feature extraction with emphasis on feature-based catchment classification (see, e.g., Burn and Boorman, 1992; Wagener et al., 2007; Sivakumar, 2008; Castellarin et al., 2011; He et al., 2011a; Ley et al., 2011; Di Prinzio et al., 2011; Sawicz et al., 2011; Ali et al., 2012; Coopersmith et al., 2012; Thomas et al., 2012; Toth, 2013; Sawicz et al., 2014; Sikorska et al., 2015; Sivakumar et al., 2015; Auerbach et al., 2016; Ley et al., 2016; Tongal and Sivakumar, 2017; Jehn et al., 2020), being viewed as (1) an effective approach to advance process perception and understanding, and (2) a necessary step for the regionalization of hydrological models (e.g., Merz and Blöschl, 2004) under a more general interpretation of hydrological similarity (i.e., similarity in specific hydrological features).

Other important concepts traditionally studied through feature extraction are those of variability and change (e.g., Montanari et al., 2013), with various trend, seasonality and temporal dependence features being among the most popular ones (e.g., Montanari et al., 1997; Grimaldi, 2004; Koutsoyiannis, 2011; Markonis and Koutsoyiannis, 2013; Mallakpour and Villarini, 2015; Prosdocimi, et al. 2015; Archfield et al., 2016; Villarini, 2016; Blöschl et al., 2017; Hall and Blöschl, 2018; Markonis et al., 2018b; Tyralis et al., 2018; Blöschl et al., 2019b; Bertola et al., 2020; Kagawa-Viviani and Giambelluca, 2020; Kelder et al., 2020; Serinaldi et al., 2020b). Both variability and change, similar to feature extraction itself, are relevant to all temporal scales (i.e., daily, monthly, seasonal, annual and inter-annual), while feature selection is usually regarded as problem-dependent (as each study or research aim may require its own set of features). For a specific problem, a representative feature set could be determined either experimentally (see, e.g., the extensive investigations by Tyralis et al., 2019c) or according to available past experience and experts' knowledge (see, e.g., the guidelines by McMillan et al., 2017).

Unarguably, (mostly) during the last decade the study of hydroclimatic variable features has rapidly progressed following the increasing release of big hydroclimatic datasets, with many diverse and important topics being advanced through large-scale feature extraction, conducted at continental-scale regions (mostly in North America and Europe) and even at the global scale. Among the most characteristic topics are those related to:

o  Temperature means (see, e.g., Papacharalampous et al., 2018; Kagawa-Viviani and Giambelluca, 2020).

o  Temperature extremes (see, e.g., Papalexiou et al., 2018b; Kagawa-Viviani and Giambelluca, 2020).

o  Precipitation means (see, e.g., Peel et al., 2002; Small et al., 2006; Papacharalampous et al., 2018; Tyralis et al., 2018).

o  Precipitation extremes (see, e.g., Papalexiou and Koutsoyiannis, 2012, 2013; Langousis et al., 2016b; Papalexiou et al., 2018a; Nerantzaki and Papalexiou, 2019; Papalexiou and Montanari, 2019; Kelder et al., 2020).

o  Streamflow means (see, e.g., Small et al., 2006; Markonis et al., 2018b; Papacharalampous et al., 2019a; Papacharalampous and Tyralis, 2020).

o  Floods and streamflow maxima (see, e.g., Villarini et al., 2011; Mallakpour and Villarini, 2015; Archfield et al., 2016; Berghuijs et al., 2016; Slater and Villarini, 2016; Villarini, 2016; Berghuijs et al., 2017; Blöschl et al., 2017; Do et al., 2017; Slater and Villarini, 2017; Steirou et al., 2017; Hall and Blöschl, 2018; Berghuijs et al., 2019a, 2019b; Bertola et al., 2020; Blöschl et al., 2019b; Iliopoulou et al., 2019; Steirou et al., 2019; Tyralis et al., 2019c; Brunner et al., 2020; Do et al., 2020; Kemter et al., 2020; Perdios and Langousis, 2020; Stein et al., 2020; see also the overviews by Hall et al., 2014; Blöschl et al., 2015; Zaghloul et al., 2020).

o  Droughts and low flows (see, e.g., Tongal et al., 2013; Van Loon et al., 2014; Hanel et al., 2018; Markonis et al., 2018a; Iliopoulou et al., 2019).

## 1.4 Aims and novelties of the present work

Here we: (1) develop a detailed framework for complete hydroclimatic variable characterization through massive feature extraction (with this massive character being the major theme of our work); (2) develop a new hydroclimatic time series clustering methodology that can also be perceived as a geographical location clustering methodology (since hydroclimatic time series correspond to geographical locations), thereby formalizing the identification of spatial patterns related to the spatial variability of the features; and (3) demonstrate the new framework and the new methodology within three global-scale applications to (a) characterize three hydroclimatic variable types (i.e., mean monthly temperature, total monthly precipitation and mean monthly river flow) and their spatial patterns, (b) inspect the statistical (dis)similarities of these variable types, (c) investigate the relationships between features, and (d) characterize the importance of the features in terms of variance explanation and in hydroclimatic time series clustering.

Our work complements the existing literature on the topic, while advancing the state-of-the-art knowledge, as summarized here below:

o  Massive extraction of a variety of features is performed for three global hydroclimatic datasets. More precisely, 59 features are computed and extracted, without particular focus on a specific category (e.g., on trend or autocorrelation features).

o  Based on these features, a new hydroclimatic time series clustering methodology is proposed. With this methodology, we aspire to exploit a larger part of the available information encompassed in monthly hydroclimatic time series than we could with existing methodologies.

o  Feature compilation is supported by past experience and experts' knowledge. The latter is largely sourced from scientific fields beyond geosciences (e.g., neuroscience, biology, biomedicine, forecasting), and has not been exploited so far in hydroclimatic and environmental contexts. Therefore, it should be encountered as a new concept for such contexts, whose usefulness is demonstrated through the present work.

o  Focus spreads to three hydroclimatic variable types (examined at the same timescale), rather than being limited to a single one, thus allowing for direct comparisons within and across hydroclimatic variables under the concept of hydrological similarity.

## 2. Data and methods

In this Section, we present our data and methods. Statistical software information is independently summarized in Appendix A.

### 2.1 Global hydroclimatic datasets

We compile a global dataset based on three large freely available temperature (Menne et al., 2018), precipitation (Peterson and Vose, 1997), and river flow (Do et al., 2018) datasets. Basic data retrieval information is provided in Appendix B. We extract complete 40-year-long mean monthly time series from 2 432 temperature stations and complete 40-year-long total monthly time series from 5 071 precipitation stations. From the entire river flow dataset, we first retrieve all the 40-year-long mean monthly river flow time series that have resulted from the aggregation of daily time series with missing up to 1% of their values (i.e., 5 849 time series). From these initially retrieved time series, we retain those that pass a visual inspection quality control (i.e., 5 601 time series that are not characterized by abrupt changes in their mean and variance nor by other irregularities that might be due to human activities). The geographical locations of the 13 104 exploited stations are presented in Figure 1.

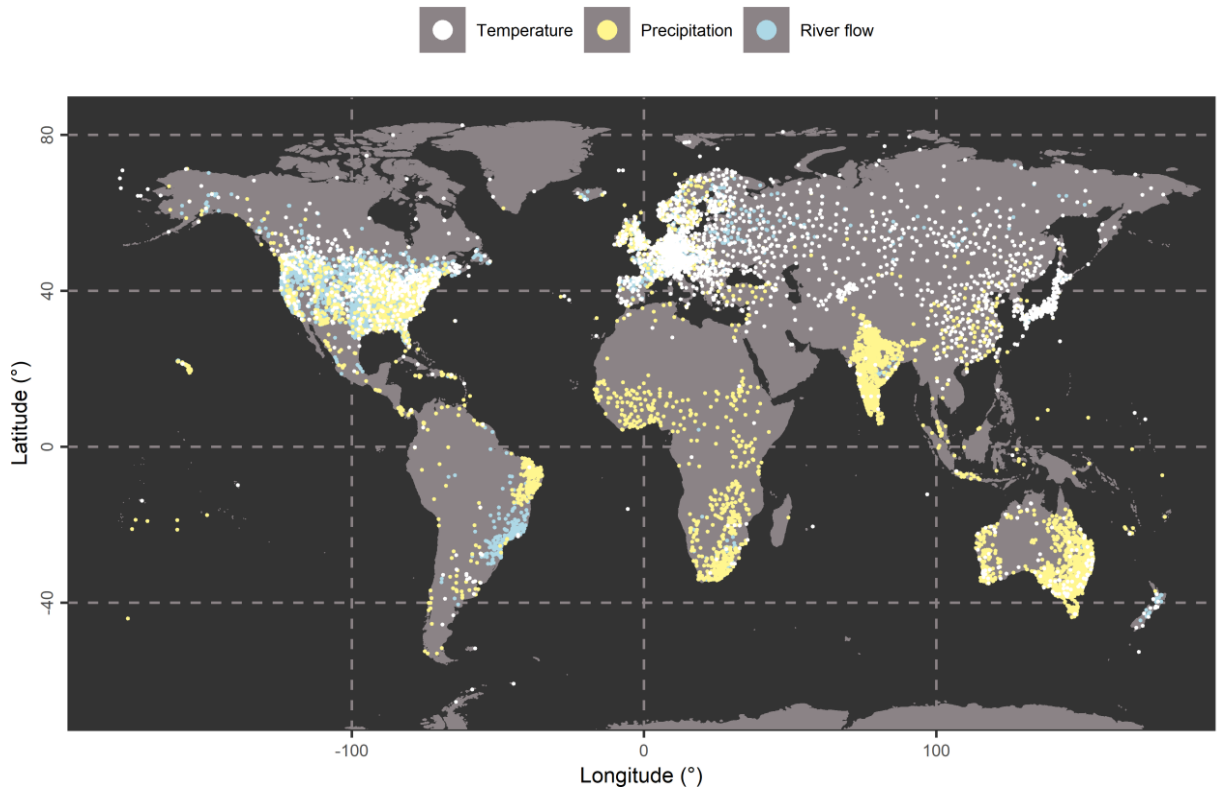

Figure 1. Geographical locations of the temperature (Menne et al., 2018), precipitation (Peterson and Vose, 1997), and river flow (Do et al., 2018) stations exploited in the study.

## 2.2 Feature description and extraction

Our methodological framework extends hydroclimatic signatures to a comprehensive set of 59 features. This feature set characterizes a wide range of the entire information content encompassed in hydroclimatic (and other geophysical) time series. A brief description of these features, adapted from Kang et al. (2017), Hyndman et al. (2020b) and Kang et al. (2020), is provided in Table 1. For their computation, we use various statistical-stochastic models, such as the sample autocorrelation function, the sample partial autocorrelation function, time series decomposition models, and the autoregressive fractionally integrated moving average (ARFIMA) model, to name a few (see Table 1 for details). Stochastic models (see, e.g., Box and Jenkins, 1970; Wei, 2006) are certainly of interest to hydrologists and geoscientists (see, e.g., Hurst, 1951; Carlson et al., 1970; Yevjevich, 1987; Hipel and McLeod, 1994; Montanari et al., 1997; Koutsoyiannis, 2002; Grimaldi, 2004; Sivakumar and Berndtsson, 2010; Veneziano and Langousis, 2010; Grimaldi et al., 2011; O'Connell et al., 2016); however, in most hydrological and geoscientific studies this interest is limited to a small number of models (which is not the case herein). Another important remark to be made, at this point, is that the delivered feature values do not depend on the scale (i.e., the magnitude) of the time series (to not be confused with the temporal resolution of the time series), although some of the features do in their original form. This holds because the time series are scaled to mean 0 and standard deviation 1 before the features are computed, and as a consequence our methodological framework allows direct comparisons within and across geophysical and environmental variables (including hydroclimatic variables).

Table 1. Features computed in the study. Their brief description is adapted mostly from the R package `tsfeatures` (Hyndman et al., 2020b) but also from the works by Kang et al. (2017, 2020) and the vignette entitled "Introduction to the `tsfeatures` package" (https://github.com/robjhyndman/tsfeatures/blob/master/vignettes/tsfeatures.Rmd), while further information can be found in Fulcher et al. (2013), Hyndman et al. (2015), and Fulcher and Jones (2017). A loose categorization of the features is provided in the Supplement (see Appendix C), specifically in Table S1 therein. The time series are scaled to mean 0 and standard deviation 1 before the features are computed.

| S/n | Abbreviation | Brief description |
|---|---|---|
| 1 | `x_acf1` | Lag-1 sample autocorrelation of the original time series. |
| 2 | `ac_9` | Lag-9 sample autocorrelation of the original time series, an autocorrelation feature from the software package `hctsa` (Fulcher and Jones, 2017; see also Fulcher et al., 2013) that is included for competency purposes (along with other autocorrelation features from the same package; see Table S1) in the software package `tsfeatures` (Hyndman et al., 2020b) and herein. |
| 3 | `x_acf10` | Sum of the squared sample autocorrelation values for the first ten lags of the original time series. |

| S/n | Abbreviation | Brief description |
|---|---|---|
| 4 | `diff1_acf1` | Lag-1 sample autocorrelation of the first-order differenced time series. |
| 5 | `diff1_acf10` | Sum of the squared sample autocorrelation values for the first ten lags of the first-order differenced time series. |
| 6 | `diff2_acf1` | Lag-1 sample autocorrelation of the second-order differenced time series. |
| 7 | `diff2_acf10` | Sum of the squared sample autocorrelation values for the first ten lags of the second-order differenced time series. |
| 8 | `seas_acf1` | Sample autocorrelation for the lag equal to twelve months. |
| 9 | `firstzero_ac` | Lag where the first zero crossing of the autocorrelation function is attained. |
| 10 | `firstmin_ac` | Lag where the first minimum of the autocorrelation function is attained. |
| 11 | `embed2_incircle_1` | Proportion of points inside a given circular boundary in a 2-dimensional embedding space. The sample autocorrelation function is computed and the lag $\tau$ for which the first zero crossing of the autocorrelation function is attained is determined. Then, a segment of the original time series $\{x_t\}$ with length $N$ equal to the length of the original time series minus $\tau$ and its lagged time series $\{z_t\}$ (with lag equal to $\tau$) are extracted. Finally, the proportion of points inside the circular boundary is equal to $\Sigma_t(I(z_t^2 + x_t^2 < \text{boundary}))/N$, where $I(\cdot)$ is the indicator function. For `embed2_incircle_1`, the parameter boundary is set to 1. |
| 12 | `embed2_incircle_2` | Proportion of points inside a given circular boundary in a 2-dimensional embedding space (see above). The parameter boundary is set to 2. |
| 13 | `trev_num` | The numerator of the `trev` function, a normalized nonlinear autocorrelation function from the software package `hctsa`. The feature is computed for the first lag of the original time series. Specifically, a segment of the original time series $\{x_t\}$ with length $N$ equal to the length of the original time series minus 1 and its lagged time series $\{z_t\}$ (with lag equal to 1) are determined. Then, `trev_num` equals to $\Sigma_t(z_t - x_t)^3/N$. |
| 14 | `motiftwo_entro3` | A feature for identifying local motifs in a binary symbolization of the time series. For its computation, coarse-graining is performed; time series values above the mean are set to 1, and those below the mean are set to 0. By using this binary symbolization of the time series, a set of counts is computed. Then, `motiftwo_entro3` is the entropy of this set of counts. |
| 15 | `walker_propcross` | A feature from the software package `hctsa` (see the function `PH_Walker` therein). A hypothetical particle (or 'walker'), moving through the time domain, is simulated. The walker starts at zero and moves in response to the values of the given time series at each point, narrowing the gap between its value and that of the given time series by 10%. More precisely, the simulated time series $\{w_t\}$ has (a) length equal to the length of the given time series, (b) its first point equal to zero and (c) its remaining points given by the following equation: $w_t = w_{t-1} + 0.1 \times (y_{t-1} - w_{t-1})$. Then, `walker_propcross` is the fraction of the length of the given (and the simulated) time series for which the given and the simulated time series get crossed in time domain. |
| 16 | `x_pacf5` | Sum of the squared sample partial autocorrelation values for the first five lags of the original time series. |
| 17 | `diff1x_pacf5` | Sum of the squared sample partial autocorrelation values for the first five lags of the first-order differenced time series. |
| 18 | `diff2x_pacf5` | Sum of the squared sample partial autocorrelation values for the first five lags of the second-order differenced time series. |
| 19 | `seas_pacf` | Sample partial autocorrelation for the lag equal to twelve months. |
| 20 | `localsimple_mean1` | Lag $\tau$ where the first zero crossing of the autocorrelation function of the residuals from a local mean prediction is attained. The prediction is obtained by using the immediate preceding value in the time series. |
| 21 | `localsimple_lfitac` | Lag $\tau$ where the first zero crossing of the autocorrelation function of the residuals from a local linear prediction is attained. More precisely, the prediction is obtained by applying a linear model, with predictors the last three values in the time series. |
| 22 | `sampen_first` | The second sample entropy of a time series, modified from Fulcher's `EN_SampEn` (Fulcher and Jones, 2017, Supplementary information), which uses a code from `PhysioNet` (Richman and Moorman, 2000). |
| 23 | `std1st_der` | Standard deviation of the first-order differenced time series. |
| 24 | `spreadrandomlocal_meantaul_50` | Bootstrap-based stationarity measure from the software package `hctsa`. First, 100 time series segments, each containing 50 consecutive points, are selected at random from the time series. Then, the feature is computed as the ensemble mean of the lags where the first zero-crossing of the autocorrelation function is attained in all segments. |
| 25 | `spreadrandomlocal_meantaul_ac2` | Bootstrap-based stationarity measure from the software package `hctsa`. First, 100 time series segments, each containing $l$ consecutive entries, where $l$ equals twice the first zero-crossing of the autocorrelation function, are selected at random from the time series. Then, the feature is computed as the ensemble mean of the first zero-crossings of the autocorrelation function in all segments. |

| S/n | Abbreviation | Brief description |
| --- | --- | --- |
| 26 | `histogram_mode_10` | Mode of a data vector using a 10-bin histogram. More precisely, a 10-bin histogram is computed for the given time series and the bin corresponding to the largest frequency value is identified. Then, `histogram_mode_10` is the center of this bin. If more than one bin are identified to correspond to the largest frequency value, then `histogram_mode_10` is the mean of the centers of these bins. |
| 27 | `outlierinclude_mdrmd` | Feature measuring the evolution of the median of a sample, as more and more outliers, located further from the mean, are included in the calculation according to a specified rule. The threshold for including data points in the analysis increases from zero to the maximum deviation, in increments of $0.01 \times \sigma$, where $\sigma$ is the standard deviation of the time series. At each threshold, proportion of time series points included and median are calculated, and outputs from the algorithm measure how these statistical quantities change as more extreme points are included in the calculation. Outliers are defined as the furthest from the mean. |
| 28 | `fluctanal_prop_r1` | Fluctuation analysis is performed. A polynomial of order 1 is fitted and the range is returned. The order of fluctuations is 2, corresponding to root mean square fluctuations. |
| 29 | `crossing_points` | The number of times a time series crosses the median. |
| 30 | `entropy` | The spectral entropy of a time series computed from a univariate normalized spectral density, estimated using an autoregressive (AR) model (see also Jung and Gibson, 2006). This feature can be used as a measure of time series "forecastability", with smaller values indicating larger forecastability (Goerg, 2013). |
| 31 | `flat_spots` | The number of flat spots in the time series, computed by dividing the sample space of a time series into ten equal-sized intervals, and computing the maximum run length within any single interval. |
| 32 | `arch_acf` | First, the original time series is pre-whitened using an AR model resulting in a new time series $\{y_t\}$. Then, `arch_acf` is calculated as the sum of the squares of the first 12 autocorrelation values of $\{y_t^2\}$. |
| 33 | `garch_acf` | First, the original time series is pre-whitened using an AR model resulting in a new time series $\{y_t\}$. Then, a GARCH(1,1) model is fitted to $\{y_t\}$ and the residuals $\{e_t\}$ are obtained. Lastly, `garch_acf` is calculated as the sum of the squares of the first 12 autocorrelation values of $\{e_t^2\}$. |
| 34 | `arch_r2` | First, the original time series is pre-whitened using an AR model resulting in a new time series $\{y_t\}$. Then, `arch_r2` is calculated as the $R^2$ value of an AR model fitted to $\{y_t^2\}$. |
| 35 | `garch_r2` | First, the original time series is pre-whitened using an AR model resulting in a new time series $\{y_t\}$. Then, a GARCH(1,1) model is fitted to $\{y_t\}$ and the residuals $\{e_t\}$ are obtained. Lastly, `garch_r2` is calculated as the $R^2$ value of an AR model fitted to $\{e_t^2\}$. |
| 36 | `alpha` | First smoothing parameter $\alpha$ of Holt-Winters' seasonal method (see, e.g., Hyndman and Athanasopoulos, 2018, Chapter 7.3), an extension of the Holt's linear trend method (see, e.g., Hyndman and Athanasopoulos, 2018, Chapter 7.2). Holt-Winters' seasonal method considers additive seasonal trend. |
| 37 | `beta` | Second smoothing parameter $\beta$ of Holt-Winters' seasonal method. |
| 38 | `gamma` | Third smoothing parameter $\gamma$ of Holt-Winters' seasonal method. |
| 39 | `lumpiness` | A feature based on 12-month-long tiled (non-overlapping) windows. First, the variances of all tiled windows are calculated. Then, `lumpiness` is the variance of the variances. |
| 40 | `stability` | A feature based on 12-month-long tiled (non-overlapping) windows. First the means of all tiled windows are calculated. Then, `stability` is the variance of the means. |
| 41 | `max_level_shift` | A feature based on 12-month-long sliding (overlapping) windows. The largest level shift (mean shift) between two consecutive windows. |
| 42 | `time_level_shift` | The time index of `max_level_shift`. |
| 43 | `max_var_shift` | A feature based on 12-month-long sliding (overlapping) windows. The largest variance shift between two consecutive windows. |
| 44 | `time_var_shift` | The time index of `max_var_shift`. |
| 45 | `max_kl_shift` | A feature based on 12-month-long sliding (overlapping) windows. The largest shift in Kulback-Leibler divergence between two consecutive windows. |
| 46 | `time_kl_shift` | The time index of `max_kl_shift`. |
| 47 | `ARCH.LM` | Feature based on the Lagrange Multiplier (LM) test of Engle (1982) for autoregressive conditional heteroscedasticity (ARCH), specifically the $R^2$ value of an autoregressive model of order 12 applied to the squared de-meaned data. |
| 48 | `nonlinearity` | Nonlinearity feature based on Teräsvirta's nonlinearity test. The feature is $10X^2/T$, where $X^2$ is the Chi-squared statistic from Teräsvirta's test, and $T$ is the length of the time series. Large values indicate nonlinearity, and values around 0 indicate linearity. |

| S/n | Abbreviation | Brief description |
|---|---|---|
| 49 | `unitroot_kpss` | The statistic of the KPSS test by Kwiatkowski et al. (1992) for testing the null hypothesis of stationarity (against the alternative hypothesis of a unit root), for linear trend and lag 1. |
| 50 | `hurst` | First, classical time series decomposition is performed using the additive model (see, e.g., Hyndman and Athanasopoulos, 2018, Chapter 6.3) according to the following equation: $x_t = S_t + T_t + R_t$. In this equation, $x_t$ denotes the data at time $t$, while $S_t$, $T_t$ and $R_t$ denote the seasonal, trend and remainder components, respectively, at time $t$. The computed feature is a measure of long-range dependence in the de-seasonalized time series (obtained by subtracting the seasonal component from the original time series), computed as 0.5 plus the maximum likelihood estimate of the fractional differencing order $d$ by Haslett and Raftery (1989; see also Maechler, 2020). |
| 51 | `trend` | First, STL decomposition (seasonal and trend decomposition using Loess; see, e.g., Hyndman and Athanasopoulos, 2018, Chapter 6.6) is applied to the original series (Hyndman and Khandakar, 2008; Hyndman et al., 2020a), according to the following equation: $x_t = S_t + T_t + R_t$. In this equation, $x_t$ denotes the data at time $t$, while $S_t$, $T_t$ and $R_t$ denote the seasonal, smoothed trend and remainder components, respectively, at time $t$. The trend strength is then measured through the following equation (Wang et al., 2006): `trend` = 1 – var($R_t$)/var($x_t - S_t$). |
| 52 | `spike` | STL decomposition (see above) is applied to the original series. The computed measure of "spikiness" is the variance of the leave-one-out variances of the remainder component $\{R_t\}$; see above. |
| 53 | `linearity` | STL decomposition (see above) is applied to the original series. The computed feature measures linearity based on the coefficients of an orthogonal quadratic regression. |
| 54 | `curvature` | STL decomposition (see above) is applied to the original series. The computed feature measures curvature based on the coefficients of an orthogonal quadratic regression. |
| 55 | `e_acf1` | STL decomposition (see above) is applied to the original series, and the sample autocorrelation function of $\{R_t\}$ (see above) is computed. Then, `e_acf1` is the lag-1 sample autocorrelation of $\{R_t\}$. |
| 56 | `e_acf10` | STL decomposition (see above) is applied to the original series, and the sample autocorrelation function of $\{R_t\}$ (see above) is computed. Then, `e_acf10` is the sum of the squared sample autocorrelation values for the first ten lags of $\{R_t\}$. |
| 57 | `seasonal_strength` | STL decomposition (see above) is applied to the original series. The seasonality strength is then measured through the following equation (Wang et al., 2006): `seasonal_strength` = 1 – var($R_t$)/var($x_t - T_t$). For the definitions of $R_t$, $x_t$ and $T_t$, see above. |
| 58 | `peak` | STL decomposition (see above) is applied to the original series. The size and location of the peaks in $\{S_t\}$ (see above) are used to compute the strength of peaks. |
| 59 | `trough` | STL decomposition (see above) is applied to the original series. The size and location of the troughs in $\{S_t\}$ (see above) are used to compute the strength of troughs. |

## 2.3  Statistical learning algorithms

In addition to the large variety of classical statistical-stochastic models utilized herein (see Section 2.2), we also apply four statistical learning (else referred to as "machine learning") algorithms (see, e.g., Hastie et al., 2009; James et al., 2013; Alpaydin, 2014). These algorithms are principal component analysis (see Section 2.3.1), linear regression (see Section 2.3.2), hierarchical clustering (see Section 2.3.3) and random forests (see Section 2.3.4). As documented in Mukhopadhyay and Wang (2020), statistical learning algorithms have their roots in nonparametric statistics. These algorithms are known to offer benefits (e.g., they can become fully automated and provide improved solutions as the dataset size increases; see also the relevant discussions in Breiman, 2001b), which are insightful for geoscientific and environmental studies (Quilty et al., 2019; Sahoo et al., 2019; Papacharalampous et al., 2019a; Tyralis et al., 2019a; Quilty and Adamowski, 2020; Rahman et al., 2020).

### 2.3.1 Principal component analysis

Principal component analysis is an unsupervised learning algorithm (i.e., an algorithm that "learns" using unlabelled data, without human "supervision") for interpretable dimensionality reduction and variance explanation characterizations, used in geosciences (Burn and Boorman, 1992; Euser et al., 2013; Jehn et al., 2020) and beyond (Abdi and Williams, 2010; Jolliffe and Cadima, 2016). For detailed literature and tutorial information on principal component analysis, the reader is referred to Abdi and Williams (2010), Bro and Smilde (2014), Shlens (2014), and Jolliffe and Cadima (2016). In summary, principal component analysis uses an input variable set to compute new variables that are linear combinations of the original ones and are obtained under the following rules: (1) The first of the new variables (also referred to as "principal components") should explain the largest possible portion of the total variance of the original (i.e., the input) data; (2) the second principal component should be orthogonal to the first principal component and explain the largest possible portion of the total variance of the original data; and (3) the remaining principal components should be computed likewise (and at the same time, they should be orthogonal to all the previous principal components). According to Jolliffe and Cadima (2016), the roots of principal component analysis trace back to Pearson (1901) and Hotelling (1933). Herein, principal component analysis is used for extracting variance explanation information.

### 2.3.2 Linear regression

Linear regression supports supervised learning (i.e., learning based on input-output examples, which can be interpreted as some sort of human "supervision"), specifically regression tasks, and is widely exploited for both statistical inference and prediction. In this work, it is used for inference and correlation analysis. Its theoretical properties are well discussed in the literature (see, e.g., Hastie et al., 2009; James et al., 2013).

### 2.3.3 Hierarchical clustering

Hierarchical clustering is a form of unsupervised learning. Once a measure of dissimilarity between (disjoint) groups of observations has been selected, the algorithm begins by assigning each data point to its own cluster and progresses by gradually joining each two most similar clusters, until a single cluster is obtained (containing all the data). At each stage, it also re-computes the distances between the clusters. Hierarchical clustering is explained in detail in Hastie et al. (2009, pp. 520–528).

### 2.3.4 Random forests

Breiman's random forests (Breiman, 2001a) belong to the family of ensemble learning algorithms (extensively reviewed, e.g., by Sagi and Rokach, 2018). They are used across a range of geoscientific and geoengineering applications (e.g., Tyralis and Papacharalampous, 2017; Addor et al., 2018; Althoff et al., 2020; Rahman et al., 2020), and big data comparisons with stochastic and/or other machine learning algorithms (see, e.g., Papacharalampous et al., 2019a, 2019b; Tyralis et al., 2020). Tyralis et al. (2019b) provide a long list of water resource engineering applications of random forests (published in 30 representative journals until 31 December 2018) together with the algorithm's theoretical summary. In short, random forests are bagging (acronym for "bootstrap aggregation"; Breiman, 1996) of classification and regression trees (CARTs; Breiman et al., 1984) with some additional degree of randomization, which aims at reducing the correlation between the trees and, consequently, the variance of the predictions (Tyralis et al., 2019b, Section 2.1.4). Their main parameter is the number of trees. They can be exploited either in supervised mode (for regression and classification) or in unsupervised mode (for clustering; see, e.g., Yan et al., 2013). In this work, they are used in both these modes, specifically for classification and clustering.

## 2.4 Feature-based time series clustering methodology

We propose a new feature-based methodology for hydroclimatic time series clustering. This methodology relies on unsupervised random forests (see Section 2.3.4) and the feature set adopted in this study (see Section 2.2), with all the computed features simply being the random forests' inputs. Together with the scale-independent nature of the delivered feature values and the large variety of the computed features (which also constitute the novelty of the methodology), the following properties of random forests contribute to making this methodology properly designed and useful for hydroclimatic time series clustering (Tyralis et al., 2019b, Section 2.8.1):

o They demonstrate increased predictive performance.

o They can handle highly correlated predictor variables.

- o They can operate successfully when interactions are present.
- o They are invariant to monotone transformations of the predictor variables.

Importantly, the proposed hydroclimatic time series methodology is applicable to different types of hydroclimatic (and environmental) data, with no need for input pre-processing via dimensionality reduction. Spatial information (i.e., the proximity of locations and spatial dependencies) and information on the time series magnitude are not considered by this methodology.

### 2.5 Global-scale application and pattern searching workflow

2.5.1 Monthly hydroclimatic time series characterizations and analyses

Hydroclimatic time series feature extraction and feature-based hydroclimatic time series analyses are conducted at monthly time scale, as outlined here below:

- o Under our feature-based approach to analyzing hydroclimatic time series data (see Section 2.2), each monthly hydroclimatic time series (see Section 2.1) is converted into a vector of 59 features, representing its corresponding time series for any further analyses. In total, 13 104 feature vectors are computed and grouped under three feature datasets, each dataset corresponding to one of the hydroclimatic variable types examined in the study. The dimensions of the resulted mean monthly temperature, total monthly precipitation and mean monthly river flow feature data matrices are 2 432 × 59, 5 071 × 59 and 5 601 × 59, respectively.

- o We summarize the information contained in the three feature datasets (see the above point) by creating histograms. On this ground, we characterize and compare the three different hydroclimatic variable types.

- o We characterize each of the three feature datasets in terms of variance explanation by conducting principal component analyses (see Section 2.3.1). As suggested in the literature (see, e.g., Bro and Smilde, 2014), pre-processing of the inputs via auto-scaling precedes the principal component analyses (because the delivered feature values are not dependent on the scale-magnitude of the time series, but range within different scales-magnitudes with regard to each other). On this ground, we search for patterns within and across the three different hydroclimatic variable types.

- o We also perform linear regression (see Section 2.3.2) and compute the linear correlations between the features for each of the feature datasets. We present the corresponding correlograms for the entire feature datasets and for those 15 features contributing the most to the first and second principal components, as these components have been previously derived through our principal component analyses (see the above point). For creating the correlograms, we additionally apply hierarchical clustering (see Section 2.3.3) to the computed correlations and test the significance of these correlations.

2.5.2 Spatial hydroclimatic patterns and hydroclimatic time series clustering

Identification of patterns related to the spatial variability of the features, as well as feature-based hydroclimatic time series clustering and hydroclimatic time series cluster characterizations are conducted at monthly timescale as outlined here below:

o We search for coherent spatial patterns characterizing the computed features by creating spatial visualizations for continental-scale regions around the globe.

o We use the computed features to apply the new time series clustering methodology (see Section 2.4), separately for each hydroclimatic variable type. In this context, random forests (see Section 2.3.4) are applied with 5 000 trees, while the number of clusters is set to five.

o We also compute variable importance (i.e., the relative significance of the variables used by the algorithm for completing a statistical learning task) and use it for ranking the features.

o Once we have obtained the clustering outcomes, we perform spatial interpolation of the clusters by applying random forests (see Section 2.3.4) in classification mode. The application is again made with 5 000 trees. Separately for each hydroclimatic variable type, we use the location-cluster information for all stations (i.e., all the available location-cluster examples) as input-output examples for training the algorithm. Once the algorithm has been trained, it can be applied to predict the cluster of an arbitrary location of the globe (including locations other than those of the stations). To allow spatial pattern extraction, we present the spatial interpolation outcomes for continental-scale regions with a large number of stations.

## 3. Results and discussion

In this section, we present the results of our global-scale analyses and investigations, and provide their interpretation in light of the study's background. A portion of the created figures is given in the Supplement (see Appendix C) for reasons of brevity herein.

**3.1 Monthly hydroclimatic time series characterizations and analyses**

3.1.1 Features of monthly hydroclimatic time series

Here, we summarize the results obtained through feature extraction. For this summary, we present and discuss a small sample of the histograms of the computed features (see Figures 2 and 3). This sample refers to ten selected features (i.e., `x_acf1`, `x_acf10`, `diff1_acf1`, `seas_acf1`, `x_pacf5`, `std1st_der`, `entropy`, `nonlinearity`, `trend` and `seasonal_strength`), for which at least one of the following conditions holds: (1) They are loosely identified (based on our intuition) as highly interpretable and/or highly relevant to the main interests spotted in the hydroclimatic literature (see the brief overview presented in Section 1.3); (2) They are objectively identified as among the top-10 important ones in explaining the variance of our three feature datasets (see Section 3.1.2), and in clustering the hydroclimatic time series (see Section 3.2.2). We choose to focus on 30 histograms for reasons of brevity only, with no intention to imply that the remaining histograms are not important, as such an implication would have been in opposition to our main premise (that by adopting a multi-faced and massive approach to hydroclimatic time series representation via feature extraction we can capture and explore the largest part of the information encoded in hydroclimatic time series). For this reason, the remaining histograms (i.e., 147 histograms) are provided in the Supplement (see Figures S1–S10 therein).

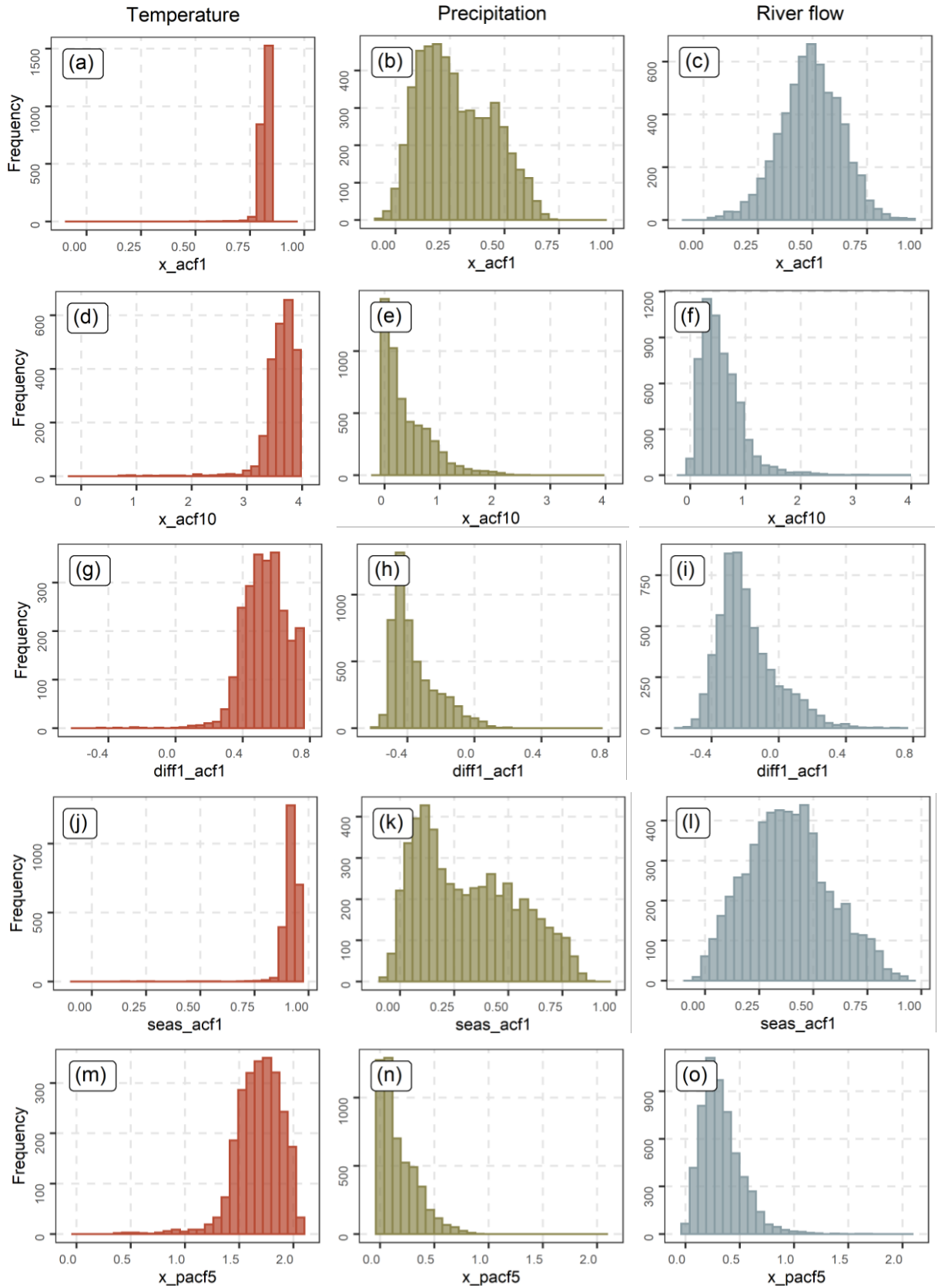

Figure 2. Histograms of the mean monthly temperature, total monthly precipitation and mean monthly river flow features (part 1). For comparison purposes, the limits of the horizontal axes have been set common for features of the same type. Outliers have been removed for `x_acf10`. The features are defined in Table 1.

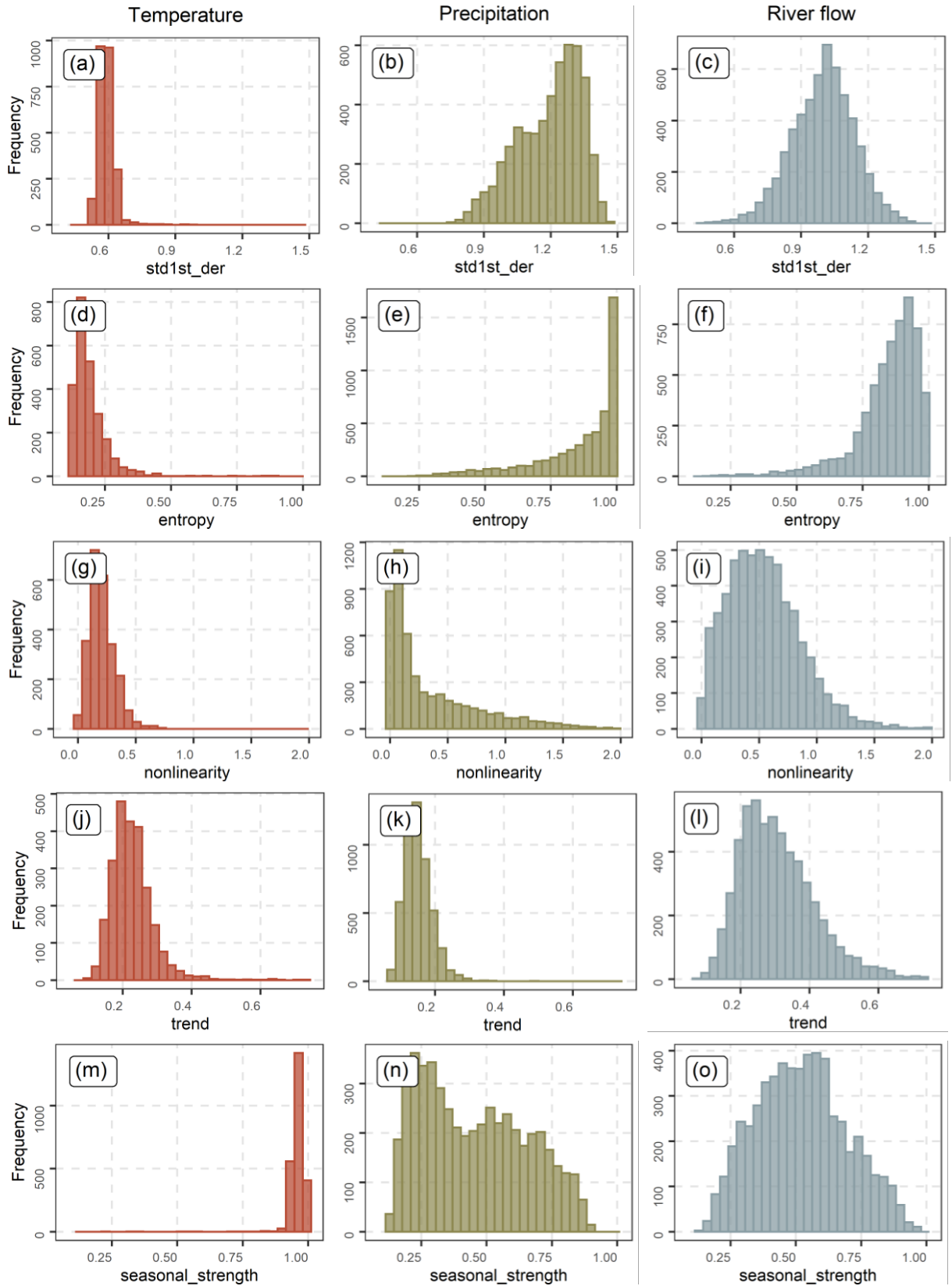

Figure 3. Histograms of the mean monthly temperature, total monthly precipitation and mean monthly river flow features (part 2). For comparison purposes, the limits of the horizontal axes have been set common for features of the same type. Outliers have been removed for `std1st_der`, `entropy`, `nonlinearity` and `trend`. The features are defined in Table 1.

To better understand what the here-provided quantitative information implies in practical terms, and to facilitate further discussions on it, it might be useful to first recall the definitions of the ten selected features (see also Table 1), as well as to discuss their utilities and relevance to hydroclimatic concepts:

o   Feature `x_acf1` (i.e., the lag-1 sample autocorrelation of the original time series) is considered a good autocorrelation measure in the hydrological literature and, thus, widely computed for interpretable characterizations of hydroclimatic variables (see e.g., Markonis et al., 2018; Papacharalampous et al., 2019a; Papacharalampous and Tyralis, 2020). It takes values from −1 to 1, and the larger its absolute value the larger the magnitude of the correlation (positive or negative, depending on the sign) between two subsequent data points in the original time series.

o   On the contrary, `x_acf10` is a new feature in hydroclimatic contexts. It is the sum of the squared sample autocorrelation values for the first ten lags of the original time series. This feature summarizes more information than `x_acf1` and is quite interpretable (yet less interpretable than `x_acf1`). It takes positive (or zero) values, and the larger its values the stronger the temporal dependence structure of the hydroclimatic variables.

o   Another feature that is scarcely computed in hydroclimatic contexts is `diff1_acf1`. This feature is defined as the lag-1 sample autocorrelation of the first-order differenced time series, and contributes –together with the remaining (partial) autocorrelation features of our framework– to better representing the temporal dependence structure of hydroclimatic variables. It takes values from −1 to 1.

o   Similar to `x_acf1`, `seas_acf1` is intuitively considered as an effective and interpretable autocorrelation measure for monthly hydroclimatic variables, as it is the sample autocorrelation value for the lag equal to twelve months. It takes values from −1 to 1.

o   In opposition to other partial autocorrelation features (e.g., the lag-1 sample partial autocorrelation of the original time series), `x_pacf5` has not been computed so far in hydroclimatic contexts to our knowledge. This feature is the sum of the squared sample partial autocorrelation values for the first five lags of the original time series; therefore, it summarizes more information than a sample partial autocorrelation value at a single lag. It takes positive (or zero) values.

- Feature `std1st_der` is the standard deviation of the first-order differenced time series and, to our knowledge, has not been computed before in hydroclimatic contexts. Since the hydroclimatic time series are scaled to mean 0 and standard deviation 1 before computing the features, the delivered `std1st_der` values allow comparisons across different variables in a similar way to the values of scaled features (e.g., `x_acf1`, `seas_acf1`, `entropy`, `trend`, `seasonal_strength`), in spite of the fact that `std1st_der` depends on the time series scale-magnitude in its original form.

- Feature `entropy` (i.e., the spectral entropy of the time series) can be used for characterizing hydroclimatic variables in terms of "forecastability" (Goerg, 2013). This feature is computed from a univariate normalized spectral density, estimated using an autoregressive (AR) model, and takes values from 0 to 1. A variety of entropy-based analyses and calibration measures can be found in the hydrological literature (see, e.g., Papalexiou and Koutsoyiannis, 2012; Pechlivanidis et al., 2014; Tongal and Sivakumar, 2017; Papacharalampous and Tyralis, 2020); nonetheless, to our knowledge the estimator used in this study is new in hydroclimatic contexts.

- Another feature that is relevant to the scientific interests identified in the hydroclimatic literature (see, e.g., Khatami, 2013; Xu et al., 2013; Fleming and Dahlke, 2014; Zhou et al., 2016) is `nonlinearity`, a feature based on Teräsvirta's nonlinearity test. This feature takes positive (or zero) values, with its large values indicating nonlinearity and its values around 0 indicating linearity.

- Hydroclimatic trends are regularly analysed (see, e.g., Khatami, 2013; Do et al., 2017; Papalexiou et al., 2018b; Khazaei et al., 2019; Bertola et al., 2020). The trend strength of the time series (i.e., `trend`) is an interpretable and dimensionless measure for investigating trends that has only been computed so far for annual (i.e., non-seasonal) river flow time series (specifically, in Papacharalampous and Tyralis, 2020) in the hydrological literature. Its usefulness arises from the fact that it allows comparisons between (a) the magnitude of the variance of the remainder component (known as the "random" component of the time series) obtained by applying STL decomposition (i.e., the variation remaining in the data after the "removal" of its seasonal and trend components; Cleveland et al., 1990) and (b) the magnitude of the time series with only its seasonal component removed, thereby informing us in relative terms on the trend's magnitude (compared to the "random" component's magnitude). This feature takes values from 0 to 1, with smaller values indicating relatively weaker trends.

o   Hydroclimatic seasonality is also of interest (see, e.g., Villarini, 2016; Hall and Blöschl, 2018). The strength of seasonality (i.e., `seasonal_strength`) is an interpretable measure for relevant investigations. Similar to `trend`, this feature provides information in relative terms on the strength of the seasonal component, obtained by applying STL decomposition, compared to the remainder component. This feature takes values from 0 to 1, with smaller values indicating relatively weaker seasonality.

As implied by our aim 3(b) (see Section 1.4), the presentation of all histograms is made in a way that allows a direct comparison between mean monthly temperature, total monthly precipitation and mean monthly river flow (see also the mean and median feature values presented in Table 2). For instance, we observe that larger values are computed for the mean monthly temperature autocorrelation and partial autocorrelation features summarized in Figure 2 (i.e., `x_acf1`, `x_acf10`, `diff1_acf1`, `seas_acf1` and `x_pacf5`) than for the total monthly precipitation and mean monthly river flow autocorrelation and partial autocorrelation features (summarized in the same figure), thereby revealing the temperature's much stronger temporal dependence characteristics. Moreover, the temporal dependence structure of mean monthly river flow is found to be somewhat more intense on average than the temporal dependence structure of total monthly precipitation (see Table 2). Similar observations are made for `seasonal_strength` (see Figure 3m–o and Table 2), indicating somewhat stronger seasonality patterns in mean monthly river flow than those in total monthly precipitation (and the strongest seasonality patterns in mean monthly temperature). The histograms of the six above-discussed features are left-skewed for mean monthly temperature (see Figures 2a,d,g,j,m and 3m), and right-skewed or not skewed for total monthly precipitation (see Figures 2b,e,h,k,n and 3n) and mean monthly river flow (see Figures 2c,f,i,l,o and 3o).

Table 2. Medians and means of ten selected mean monthly temperature, total monthly precipitation and mean monthly river flow features. The features are defined in Table 1.

| Hydroclimatic time series feature | Corresponding histograms | Summary statistic | Mean monthly temperature | Total monthly precipitation | Mean monthly river flow |
|---|---|---|---|---|---|
| x_acf1 | Figure 2a–c | Median | 0.825 | 0.251 | 0.492 |
|  |  | Mean | 0.822 | 0.278 | 0.491 |
| x_acf10 | Figure 2d–f | Median | 3.666 | 0.241 | 0.484 |
|  |  | Mean | 3.614 | 0.405 | 0.571 |
| diff1_acf1 | Figure 2g–i | Median | 0.538 | −0.403 | −0.226 |
|  |  | Mean | 0.534 | −0.357 | −0.187 |
| seas_acf1 | Figure 2j–l | Median | 0.923 | 0.281 | 0.377 |
|  |  | Mean | 0.919 | 0.316 | 0.385 |
| x_pacf5 | Figure 2m–o | Median | 1.698 | 0.112 | 0.297 |
|  |  | Mean | 1.683 | 0.172 | 0.331 |
| std1st_der | Figure 3a–c | Median | 0.587 | 1.224 | 1.007 |
|  |  | Mean | 0.591 | 1.193 | 0.998 |
| entropy | Figure 3d–f | Median | 0.174 | 0.924 | 0.875 |
|  |  | Mean | 0.191 | 0.860 | 0.848 |
| nonlinearity | Figure 3g–i | Median | 0.190 | 0.167 | 0.514 |
|  |  | Mean | 0.208 | 0.349 | 0.550 |
| trend | Figure 3j–l | Median | 0.220 | 0.145 | 0.293 |
|  |  | Mean | 0.227 | 0.150 | 0.311 |
| seasonal_strength | Figure 3m–o | Median | 0.964 | 0.432 | 0.517 |
|  |  | Mean | 0.960 | 0.453 | 0.522 |

On the other hand, the exact opposite holds for std1st_der and entropy, with their computed values being smaller for mean monthly temperature (see Figure 3a,d) than for total monthly precipitation (see Figure 3b,e) and mean monthly river flow (see Figure 3c,f), and the corresponding histograms being right-skewed for the former hydroclimatic variable type and left-skewed or approximately non-skewed for the latter two hydroclimatic variable types. Furthermore, larger std1st_der values are computed on average for total monthly precipitation than for mean monthly river flow, indicating somewhat stronger temporal variation in the former. Also notably, similar mean entropy values are computed for total monthly precipitation and for mean monthly river flow, indicating a similar degree of "forecastability" between them. Lastly, for all hydroclimatic variable types, the histograms of nonlinearity and trend are right-skewed (see Figure 3g–l), with the nonlinearity values on average smallest for mean monthly temperature and largest for mean monthly river flow (see Table 2), and the trend values on average smallest for total monthly precipitation and largest for mean monthly river flow (but in general indicating weak trend strength on average, although some large trend values are also computed, mostly for river flow).

### 3.1.2 Characterizations in terms of variance explanation

Here, we focus on the most representative characterizations in terms of variance explanation for the three feature datasets (to better understand these datasets, and the similarities and differences across the three examined hydroclimatic variable types). These characterizations are summarized in Figure 4, while a more detailed presentation is given in the Supplement (see Figures S11–S17 therein). For the below provided and discussed information, it might be useful to recall that (1) each principal component is a linear combination of our features, and (2) the total variance is collectively explained by all principal components (with the contributions of these components in variance explanation being smaller as we move from the first one to the last one); see again Section 2.3.1 for a short summary of principal component analysis.

For our three feature datasets, the first three principal components explain ~50% of the total variance (see Figure S11). The first principal component explains 30.5% of the total variance of the mean monthly temperature feature dataset (see Figure S12), while the respective percentages for the total monthly precipitation and mean monthly river flow feature datasets are 34.1% (Figure S13) and 24.1% (Figure S14). For mean monthly temperature, 18 features contribute by ~3–5% to the first principal component (maximum contribution computed for `seas_acf1`), with the remaining features contributing less to this component and more to the remaining ones (see Figure 4). The numbers of features contributing more than by 3% to the first principal component are 15 and 17 for total monthly precipitation and mean monthly river flow, respectively. The maximum contributions are ~4.5% for total monthly precipitation and 6% for mean monthly river flow, and the features exhibiting these contributions are `seasonal_strength` and `x_pacf5`, respectively. Interestingly, several feature types are among the top contributors to the first principal component for all three (or at least for two) main hydroclimatic variable types (e.g., `x_acf1`, `x_acf10`, `diff1_acf1`, `seas_acf1`, `diff2x_pacf5`, `x_pacf5`, `std1st_der`, `entropy` and `seasonal_strength`; see Figure 4a). On the contrary, the top contributing features to the second principal component are distinct for each main type of hydroclimatic variables (see Figure 4b). In the mean monthly temperature, total monthly precipitation and mean monthly river flow feature datasets, the second principal component explains 10.9%, 9.4% and 14.6%, respectively, of the total variance.

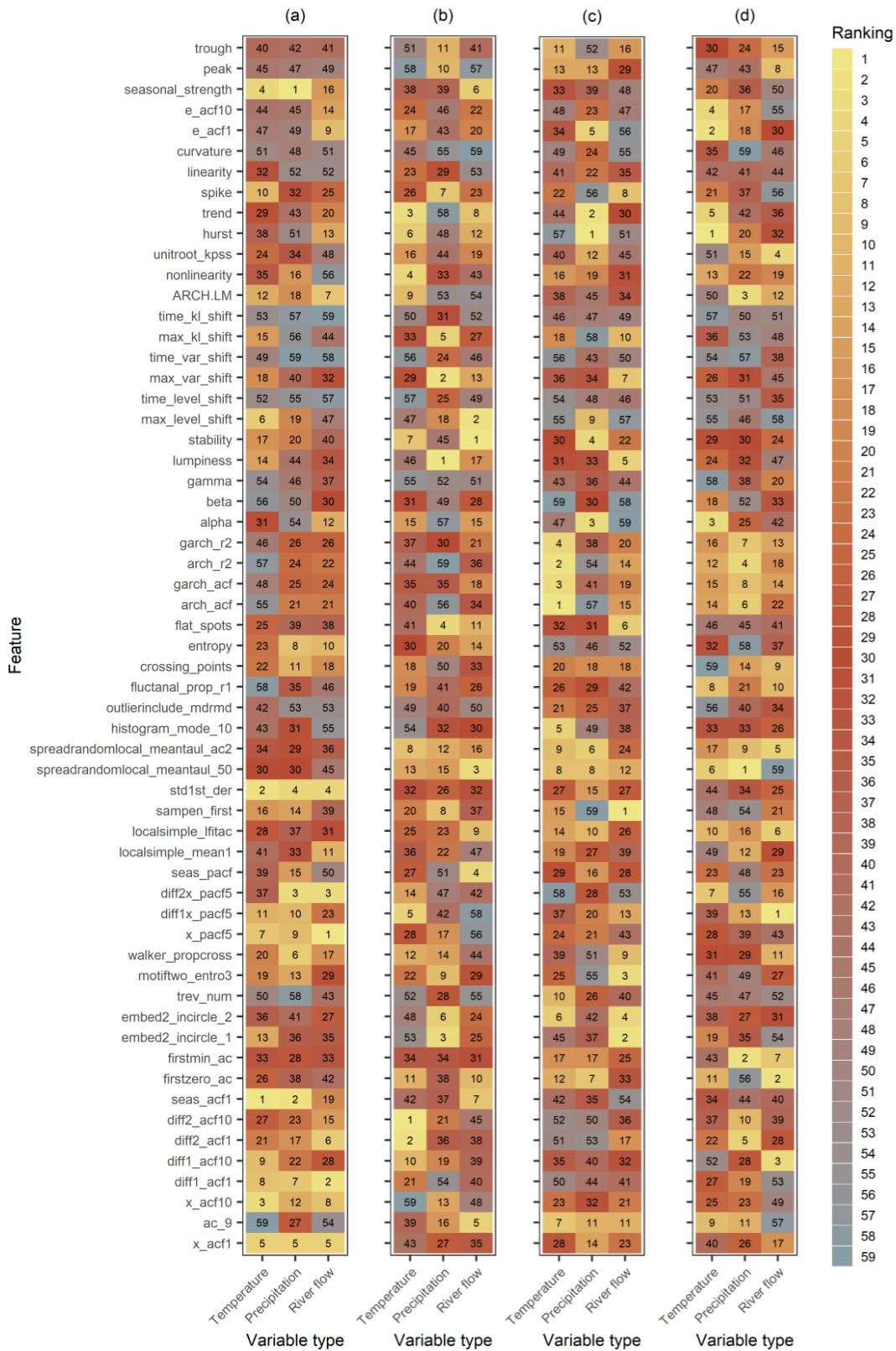

Figure 4. Rankings of the features according to their contributions to the (a) first, (b) second, (c) third, and (d) fourth principal components. (Each principal component is a linear combination of the features). The smaller a feature's ranking, the larger its contribution to a specific principal component. The contributions of all features to the first and second principal components can be found in Figures S15–S17 of the Supplement. The features are defined in Table 1.

### 3.1.3 Feature relationship characterizations and analyses

To better understand the three feature datasets obtained through massive feature extraction, and to extract further empirical evidence on the statistical similarities (and differences) characterizing the three examined hydroclimatic variable types, we examine the relationships between the features. Hierarchically clustered correlograms, created for the entire mean monthly temperature, total monthly precipitation and mean monthly river flow feature datasets, are provided in the Supplement (Figures S18–S20). Here, we focus on the top-15 contributing features to the first and second principal components for each hydroclimatic variable type (as identified in Figure 4) by presenting and discussing the hierarchically clustered correlograms of Figures 5–7. Half of these correlograms (see Figures 5a, 6a and 7a) depict medium to large (positive or negative) linear correlations, while the other half (see Figures 5b, 6b and 7b) depict linear correlations of various magnitudes (including statistically insignificant values).

As derived mainly from the former half, some features are highly correlated with each other (in terms of Pearson's correlation) for all three hydroclimatic variable types, suggesting intense feature relationships and some sort of statistical similarity across the examined hydroclimatic variable types. Among the most characteristic examples of such features are the following autocorrelation and partial autocorrelation ones: `x_acf1`, `x_acf10`, `diff1_acf1`, `diff1_acf10`, `seas_acf1`, `x_pacf5`, `diff1x_pacf5` and `seas_pacf`. These features are also found to be (mostly) highly correlated with `std1st_der`, `ARCH.LM` and `seasonal_strength`, to name a few relevant examples. Other features whose relationships are found to be intense are `entropy` and `seasonal_strength` with correlations −0.91 and −0.87 for total monthly precipitation and mean monthly river flow, respectively, and `hurst` and `trend` with correlations 0.88 and 0.86 for mean monthly temperature and mean monthly river flow, respectively. These two latter features are also found to be highly correlated with `alpha`. Statistical differences are also identified between the examined variable types. For instance, the relationships between `entropy` and several autocorrelation features are very intense for the total monthly precipitation dataset, and medium for the mean monthly temperature and mean monthly river flow datasets.

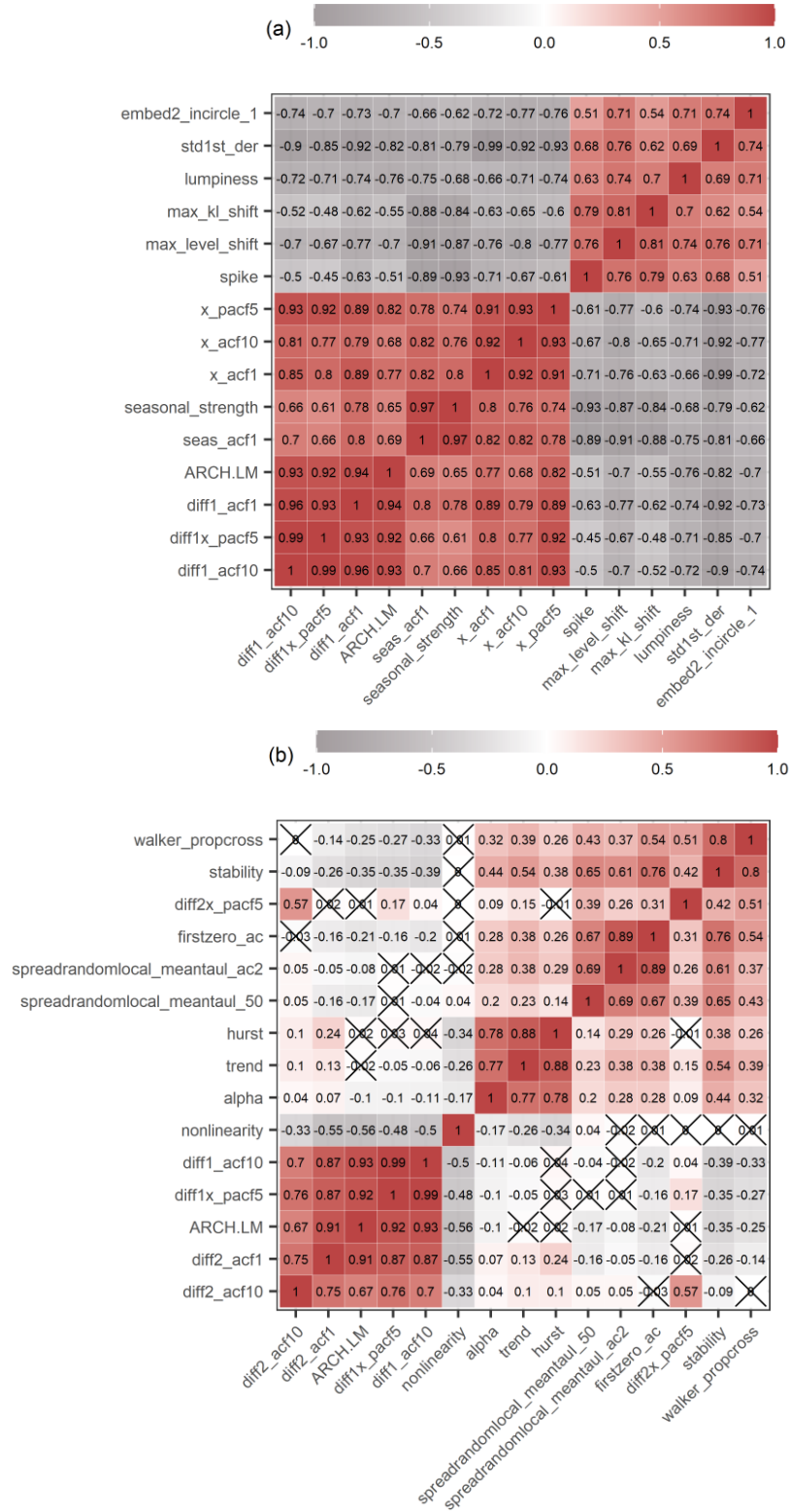

Figure 5. Linear correlations (a) between the fifteen mean monthly temperature features contributing the most to the first principal component, and (b) between the fifteen mean monthly temperature features contributing the most to the second principal component. The contributions of the mean monthly temperature features to the first and second principal components are presented in the Supplement (see Figure S15). The correlograms are hierarchically clustered. Statistically insignificant correlations are marked with crosses. The features are defined in Table 1.

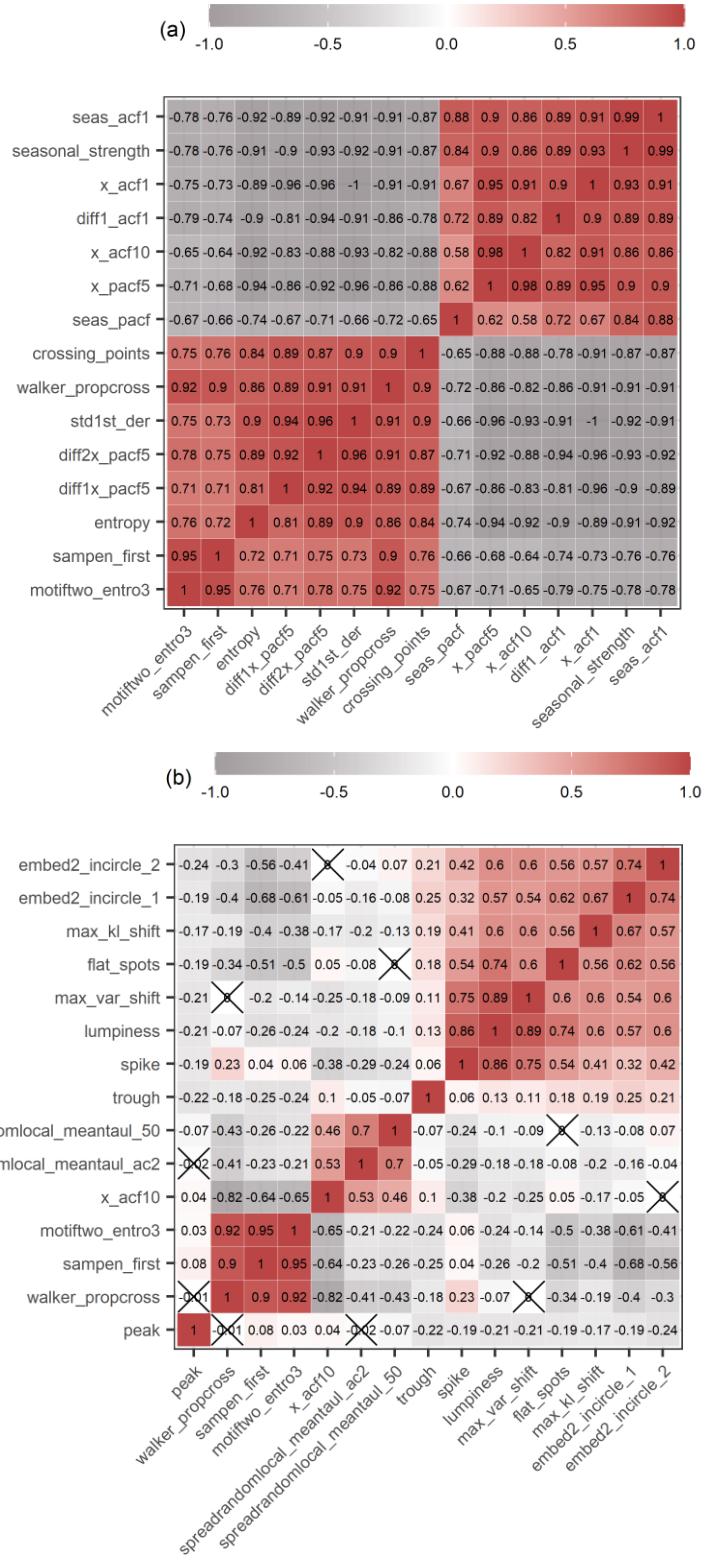

Figure 6. Linear correlations (a) between the fifteen total monthly precipitation features contributing the most to the first principal component, and (b) between the fifteen total monthly precipitation features contributing the most to the second principal component. The contributions of the total monthly precipitation features to the first and second principal components are presented in the Supplement (see Figure S16). The correlograms are hierarchically clustered. Statistically insignificant correlations are marked with crosses. The features are defined in Table 1.

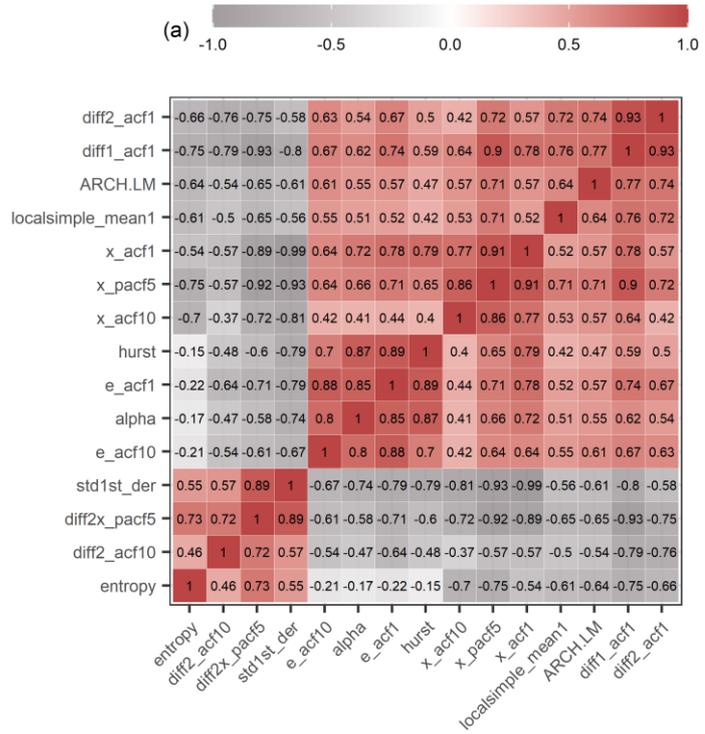

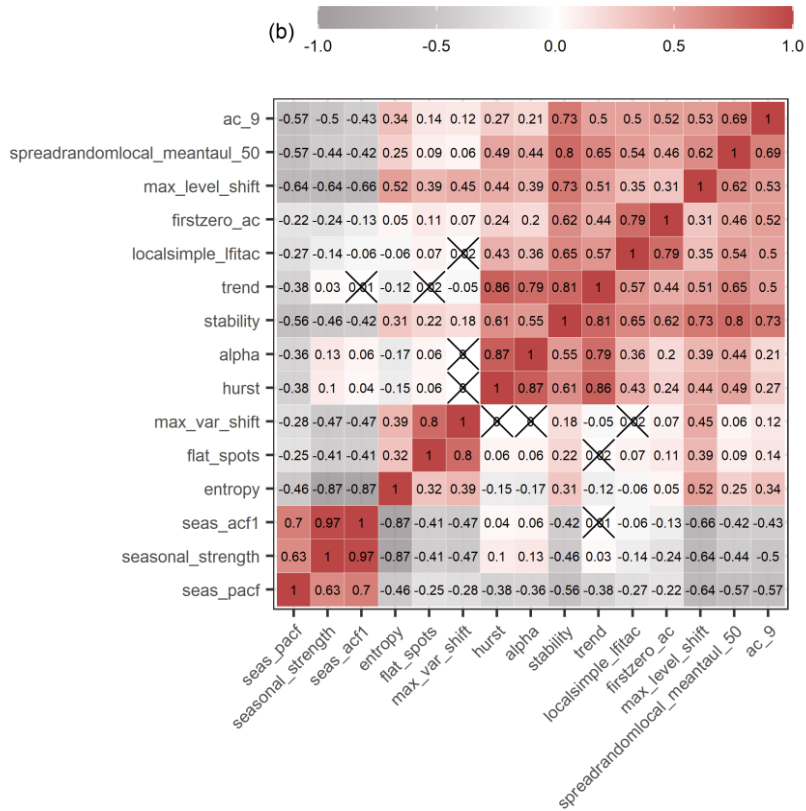

Figure 7. Linear correlations (a) between the fifteen mean monthly river flow features contributing the most to the first principal component, and between (b) the fifteen mean monthly river flow features contributing the most to the second principal component. The contributions of the mean monthly river flow features to the first and second principal components are presented in the Supplement (see Figure S17). The correlograms are hierarchically clustered. Statistically insignificant correlations are marked with crosses. The features are defined in Table 1.

## 3.2 Spatial hydroclimatic patterns and hydroclimatic time series clustering

3.2.1 Spatial variability of hydroclimatic time series features

To facilitate the detection of spatial patterns, we produce continental-scale spatial visualizations for several regions around the globe (see Figure S21 in the Supplement and Figures 8–12). These spatial visualizations refer to selected mean monthly temperature, total monthly precipitation and mean monthly river flow features. We select only four features for reasons of brevity. Specifically, we select `x_acf1`, `entropy`, `trend` and `seasonal_strength` due to their high interpretability (see also Table 1 and Section 3.1.1) and relevance to the main themes identified in the hydroclimatic literature (see Sections 1.3 and 3.1.1). Notably, the selection of `x_acf1`, `entropy` and `seasonal_strength` is also supported by their importance in terms of variance explanation (see Section 3.1.2) and in hydroclimatic time series clustering (see Section 3.2.2). Their latter importance could be helpful in inspecting and interpreting the clustering outcomes presented in Section 3.2.2.

Interestingly, total monthly precipitation across the Indian subcontinent and Australia exhibits small `trend` (see Figures 8c and 9c), without any particular patterns being noticed, except for the fact that the `trend` values are slightly larger for Australia than for India. By recalling the definition of this dimensionless feature (see Table 1), we interpret this result as follows: The variance of the remainder component obtained by applying STL decomposition is comparably large to the variance of the deseasonalized total monthly precipitation time series (i.e., the total monthly precipitation time series with only its seasonal component removed). This suggests that the trend component is relatively weak (compared to the remainder component, which can be interpreted as the "random" component of the time series; see also Section 3.1.1).

Furthermore, the other three total monthly precipitation features under study (see Figures 8a,b,d and 9a,b,d) suggest some degree of spatial coherence in these two regions. Specifically, both `x_acf1` and `seasonal_strength` are notably higher for the West Coast of India (region exhibiting tropical monsoon climate) than for the inner Indian subcontinent and the East Cost of India for the same latitudes (respectively suggesting stronger autocorrelation structure and stronger seasonality patterns for the West Coast compared to these latter regions). Moreover, `entropy` is remarkably lower for the same tropical monsoon region (suggesting larger degree of "forecastability" for total monthly precipitation variables across this region; Goerg 2013). Analogous spatial patterns can also be extracted for other regions in the Indian subcontinent and Australia; nonetheless, perhaps the most notable findings are those related with the large differences (in terms of features' magnitude) between the Indian subcontinent and Australia. For instance, it is found that `entropy` (mostly) takes much larger values across Australia than it takes across India and its neighbouring regions (suggesting that the former continental-scale region is characterized by smaller "forecastability" than the latter). Total monthly precipitation in Australia is also characterized by a less pronounced autocorrelation structure and by less intense seasonal patterns than total monthly precipitation in the Indian subcontinent (as suggested by the smaller `x_acf1` and `seasonal_strength` values computed for Australia). Also notably, Figures 8 and 9 show high (positive or negative) correlation between `x_acf1`, `entropy` and `seasonal_strength`, thereby suggesting intense relationships between them (and serving as graphical illustrations of the findings already presented and discussed in Section 3.1.3).

Distinct spatial patterns are also observed for mean monthly river flow. Since most river flow stations are from North America and Europe, Figures 10–12 are devoted to these two continental-scale regions. The (medium or) high (positive or negative) correlation between `x_acf1`, `entropy` and `seasonal_strength` is also evident for the mean monthly river flow processes of North America (see Figures 10 and 11). For the same continental-scale region, distinct sub-regions can be identified with similar feature values. For instance, central North America is characterized by smaller values of `x_acf1` and `seasonal_strength` with respect to its neighbouring sub-regions and notably larger `entropy` values. For Europe, patterns are less evident, but not ignorable. For instance, mean monthly river flow in the Alps exhibit stronger seasonality and autocorrelation patterns compared to their adjacent regions (as indicated by larger `seasonal_strength` and `x_acf1` values, respectively), and larger "forecastability" (as indicated by smaller `entropy` values). Lastly, we observe several large values for `trend`, rather scattered across North America and Europe.

Since we do not identify any patterns for mean monthly temperature, its corresponding results are only presented in the Supplement (see Figure S21 therein). As shown, `x_acf1` and `seasonal_strength` are high for all stations in and around Europe, while `entropy` and `trend` are quite low. The same applies to other regions; for justification, see the ranges of the computed features in Figures 2 and 3.

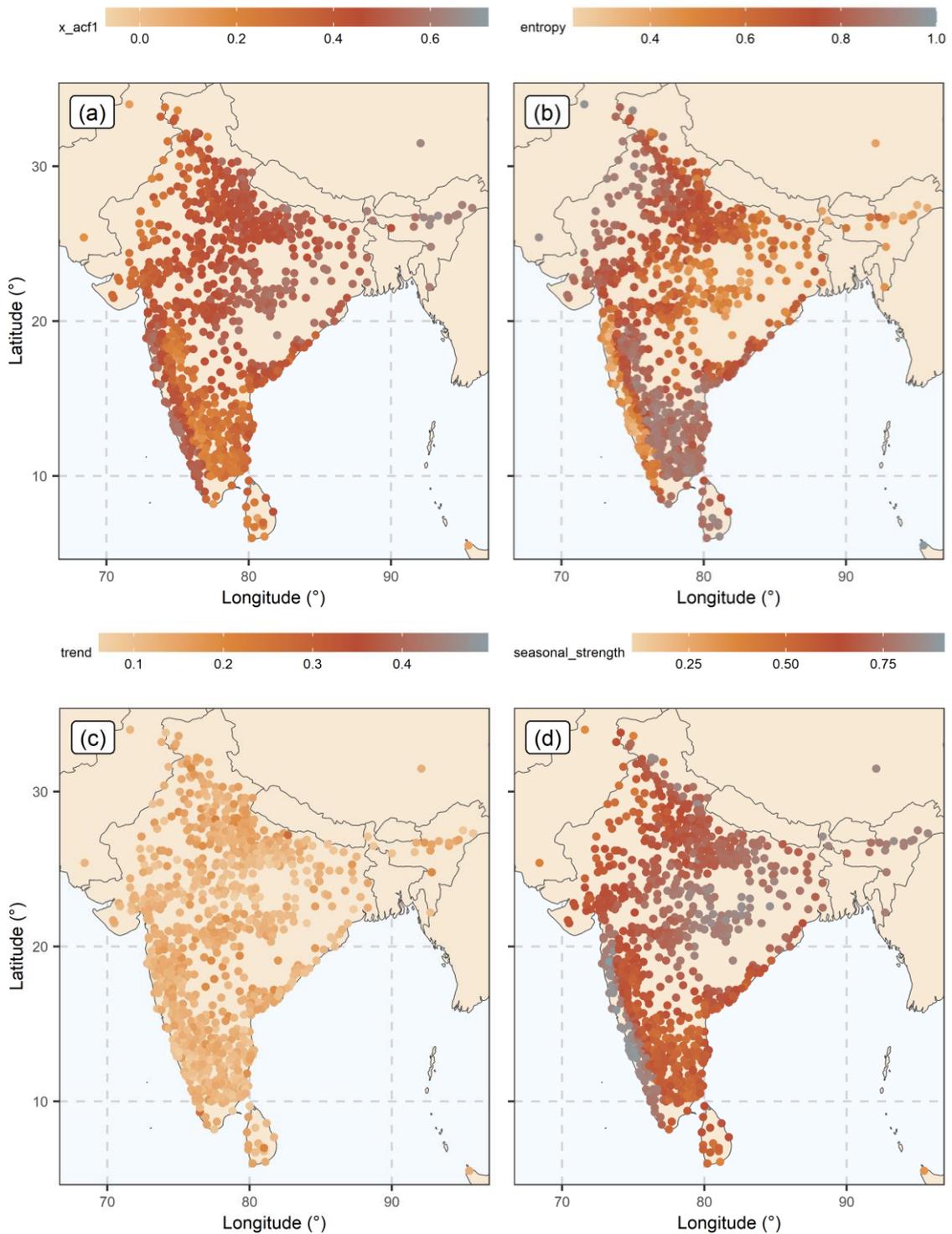

Figure 8. Selected total monthly precipitation features across the Indian subcontinent. The presented features are: (a) `x_acf1`, (b) `entropy`, (c) `trend` and (d) `seasonal_strength`. For their definitions, see Table 1. The legend bar limits are set based on the minimum and maximum values of each total monthly precipitation feature computed at the global scale.

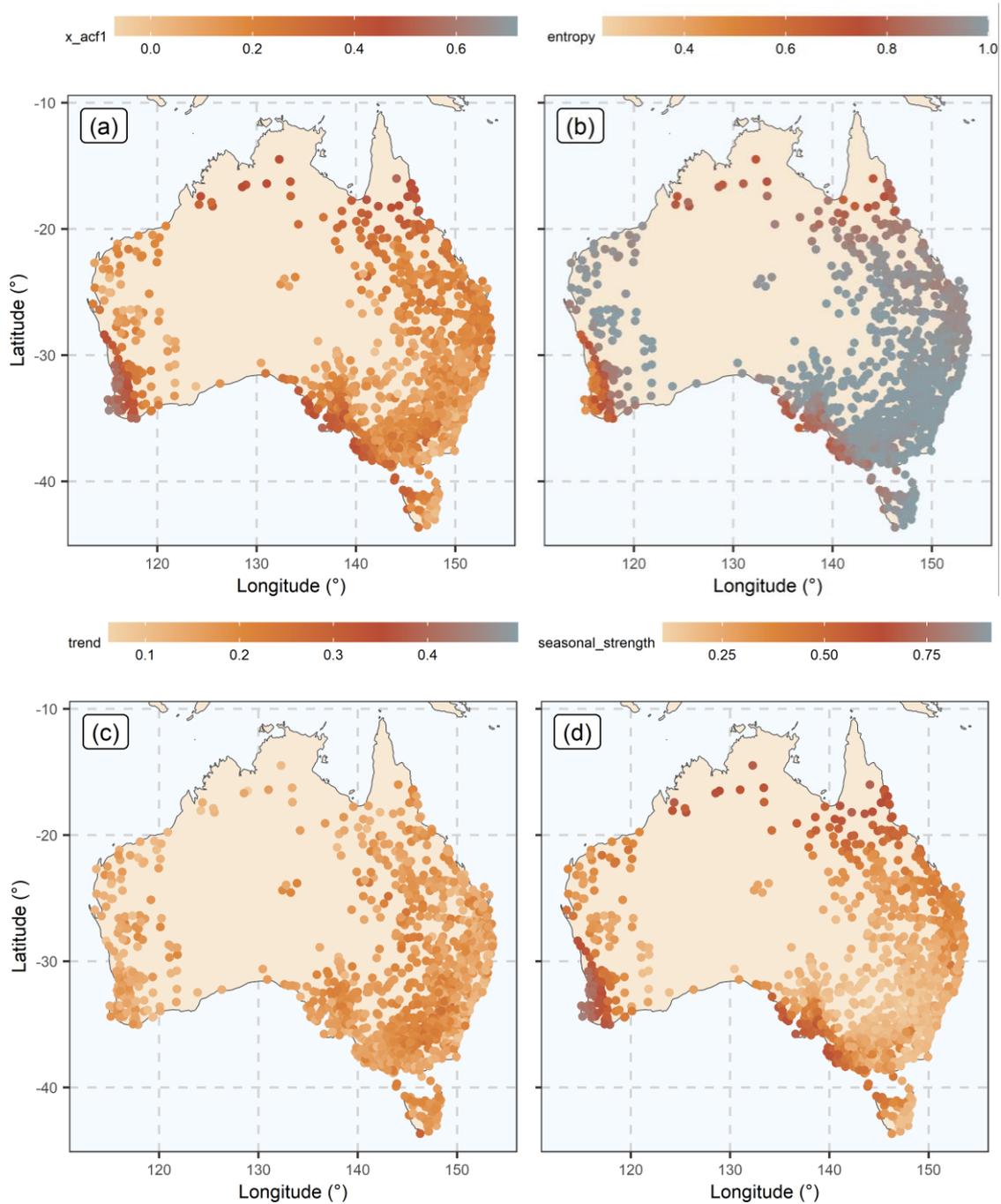

Figure 9. Selected total monthly precipitation features across Australia. The presented features are: (a) `x_acf1`, (b) `entropy`, (c) `trend` and (d) `seasonal_strength`. For their definitions, see Table 1. The legend bar limits are set based on the minimum and maximum values of each total monthly precipitation feature computed at the global scale.

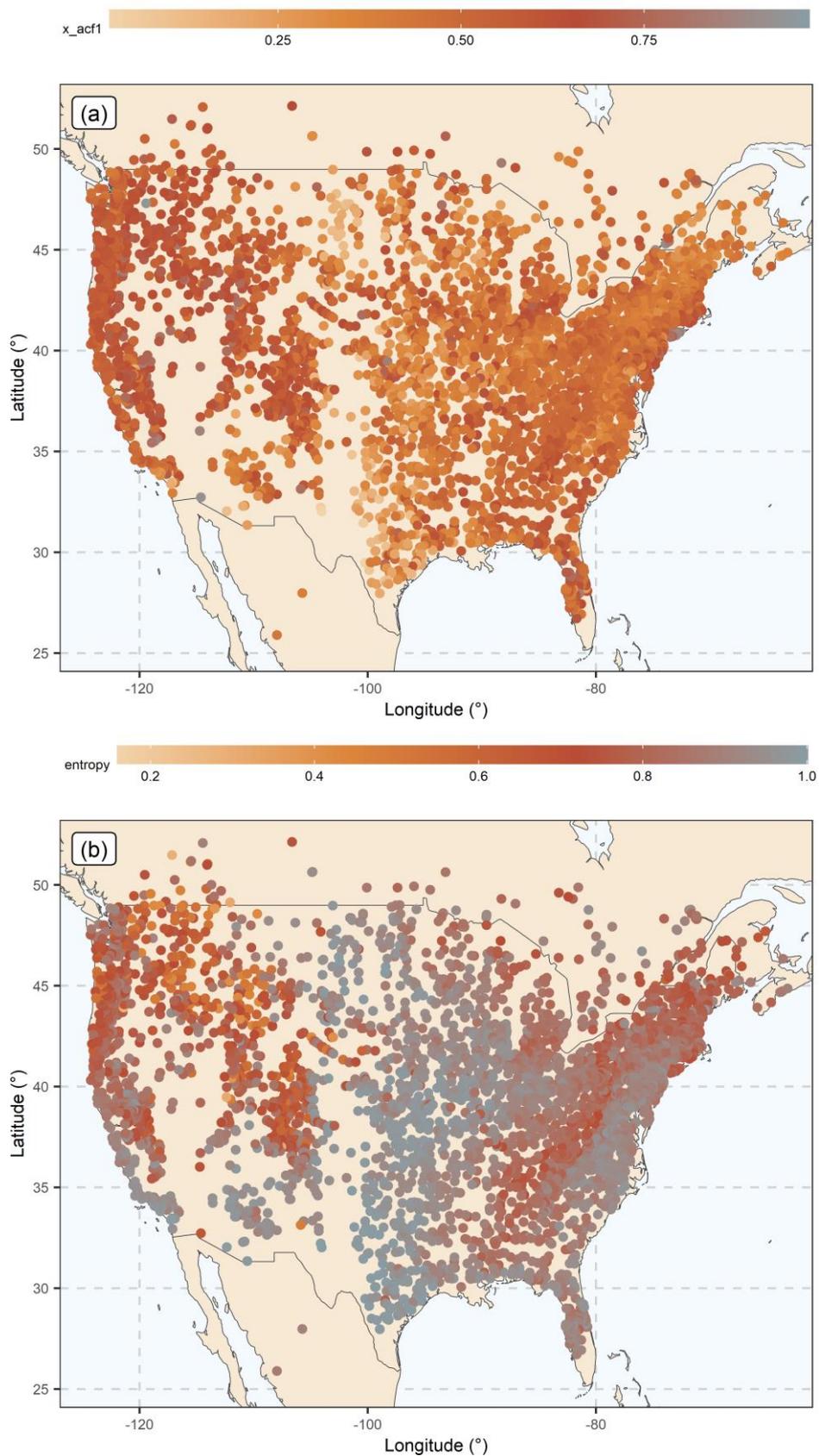

Figure 10. Selected mean monthly river flow features across North America (part 1). The presented features are: (a) `x_acf1` and (b) `entropy`. For their definitions, see Table 1. The legend bar limits are set based on the minimum and maximum values of each mean monthly river flow feature computed at the global scale.

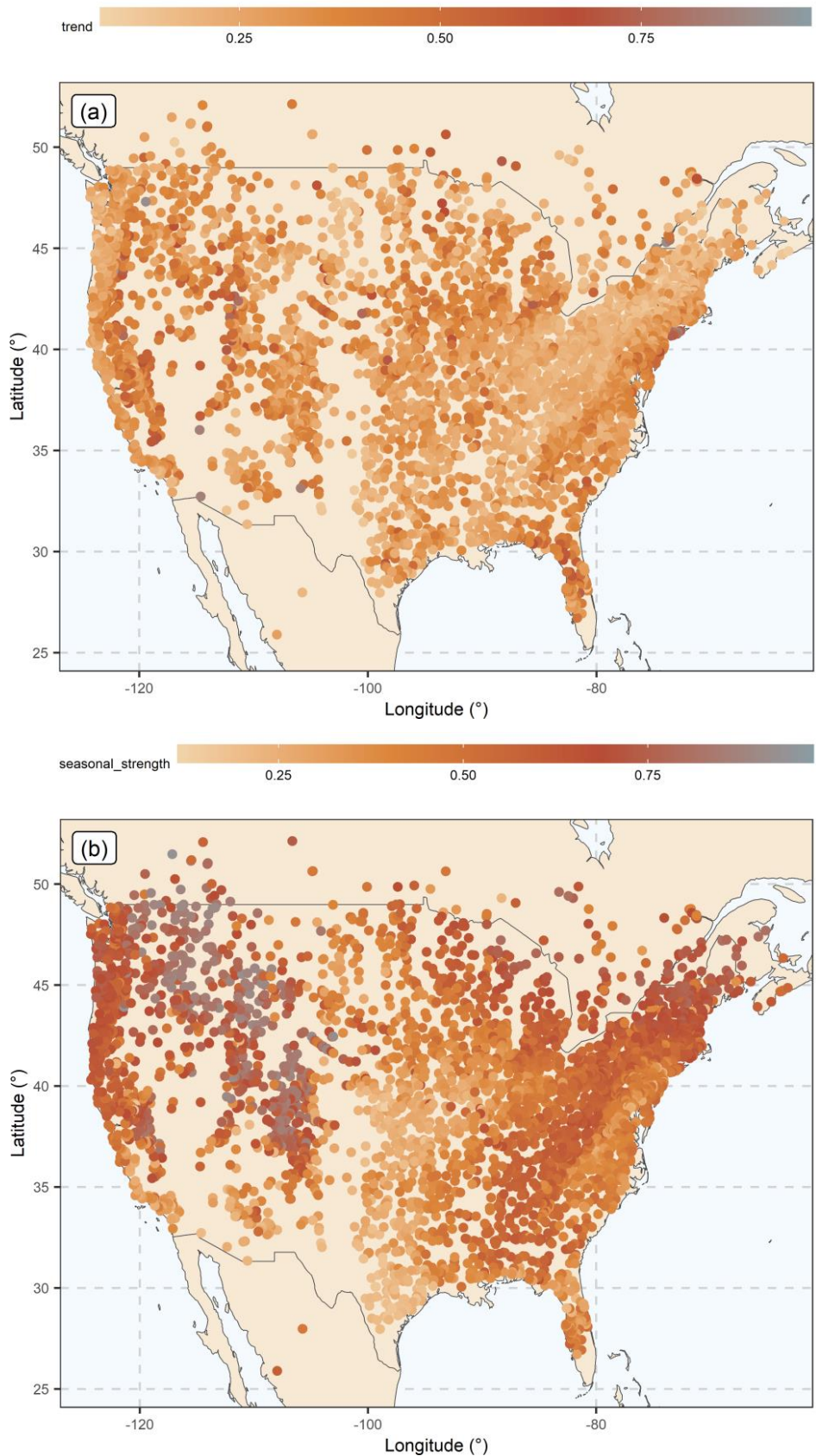

Figure 11. Selected mean monthly river flow features across North America (part 2). The presented features are: (a) `trend` and (b) `seasonal_strength`. For their definitions, see Table 1. The legend bar limits are set based on the minimum and maximum values of each mean monthly river flow feature computed at the global scale.

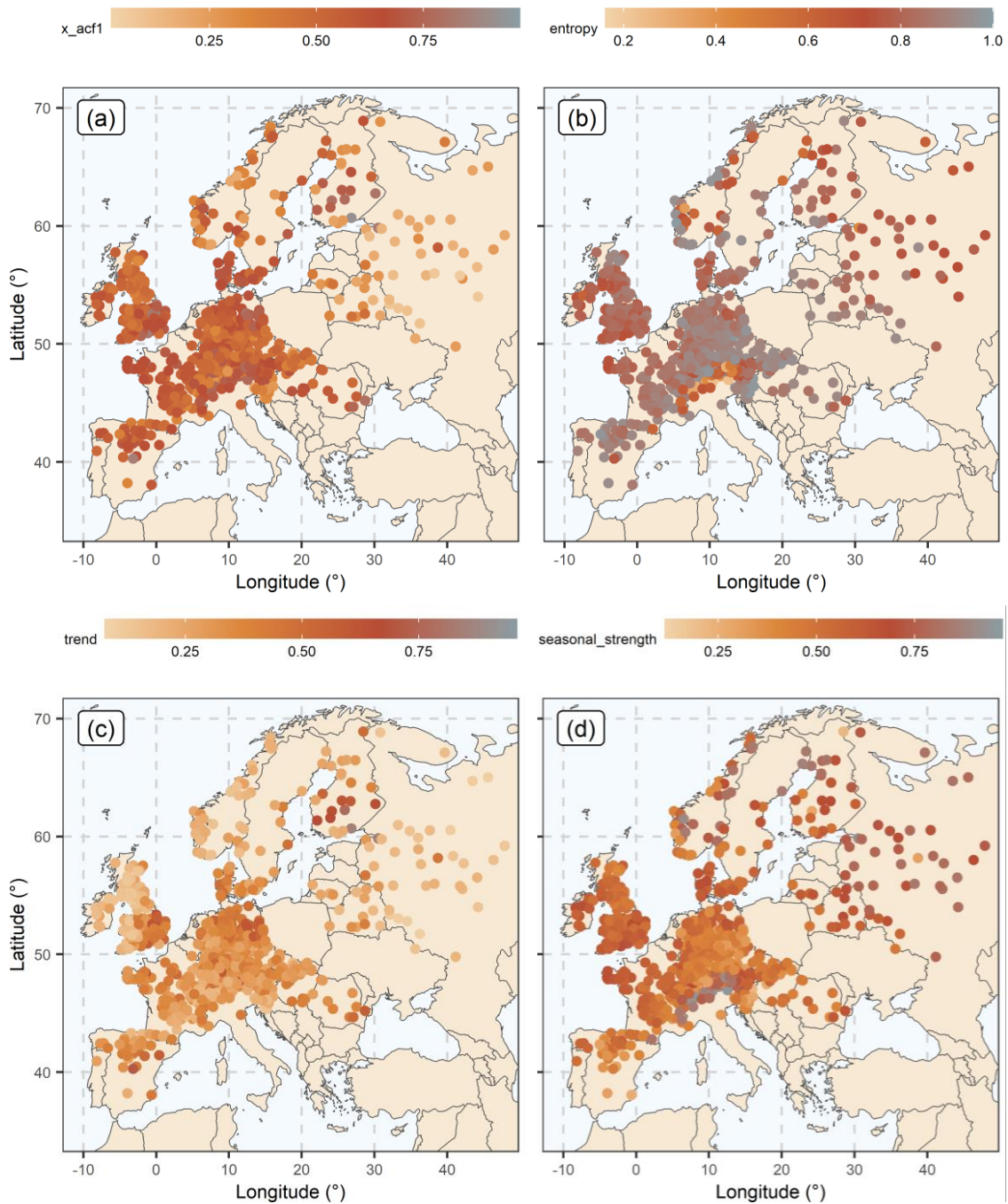

Figure 12. Selected mean monthly river flow features across Europe. The presented features are: (a) `x_acf1`, (b) `entropy`, (c) `trend` and (d) `seasonal_strength`. For their definitions, see Table 1. The legend bar limits are set based on the minimum and maximum values of each mean monthly river flow feature computed at the global scale.

3.2.2 Feature-based hydroclimatic time series clustering

Here, we present the clustering outcomes for the hydroclimatic time series examined in this work and, by extension, for the geographical locations in which these hydroclimatic time series have been observed. The presentation is made across the globe (Figure 13; see also Figure S22 in the Supplement for the percentages of the time series attributed to each cluster for each hydroclimatic variable type) and across selected continental-scale regions (Figures 14–16), with T1, T2, T3, T4 and T5 simply signifying the five clusters obtained for mean monthly temperature and similar notations being used for the five clusters obtained for total monthly precipitation (i.e., P1, P2, P3, P4 and P5), as well as for the five clusters obtained for mean monthly river flow (i.e., R1, R2, R3, R4 and R5). For the selected continental-scale regions, we also present the categorical spatial interpolation outcomes to ease the identification of spatial patterns. Regarding these spatial interpolations, it might be relevant to note that the emerging shapes are somewhat different from the shapes that we would have obtained with linear spatial interpolation, as already expected (due to the internal mechanism of the random forest algorithm). Spatial interpolation outcomes are especially meaningful for regions with high density of stations, with North America being one of the most characteristic examples of such regions for its river flow stations (see Figure 16).

The spatial coherence of the clusters is evident in Figures 13–16. For instance, the largest part of the mean monthly temperature time series observed in Europe (mostly those observed in northern regions) is attributed to the same cluster, specifically to cluster T1 (see Figure 14a), while East Asian temperature is mostly attributed to clusters T4 and T5. Distinct spatial patterns are also identified for total monthly precipitation. The Indian subcontinent is an interesting example of this hydroclimatic variable type, already examined in terms of `x_acf1`, `entropy`, `trend` and `seasonal_strength` (see Section 3.2.1, specifically Figure 8 therein). The spatially coherent patterns observed for this continental-scale region under our hydroclimatic time series clustering approach (see Figure 15b) are consistent with those previously identified (based on Figure 8), with the total monthly precipitation time series observed in the West Coast of India being attributed to a different cluster (specifically, to cluster P5) from those observed in the inner and eastern Indian subcontinent for the same latitudes (i.e., clusters P3 and P4). Lastly, distinct spatial patterns are also found for mean monthly streamflow in North America with latitudes up to approximately 50° (see Figure 16), with those time series originating from its western and northern parts belonging to cluster R1 (except for some originating from neighbouring stations that belong to cluster R2), while the time series originating from its middle and eastern parts (mostly) belonging to clusters R3 and R5, respectively.

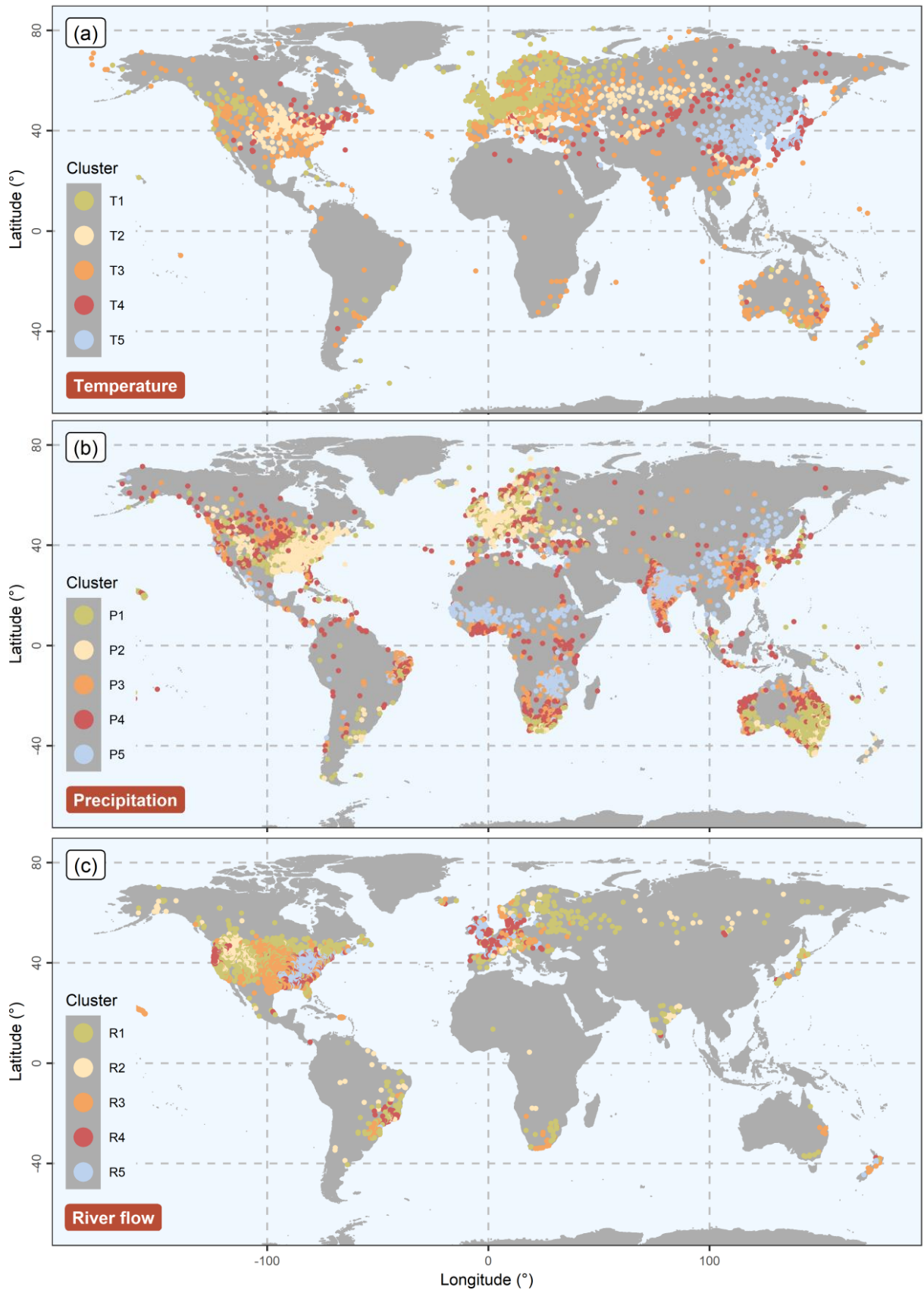

Figure 13. Clusters of the (a) mean monthly temperature, (b) total monthly precipitation and (c) mean monthly river flow time series.

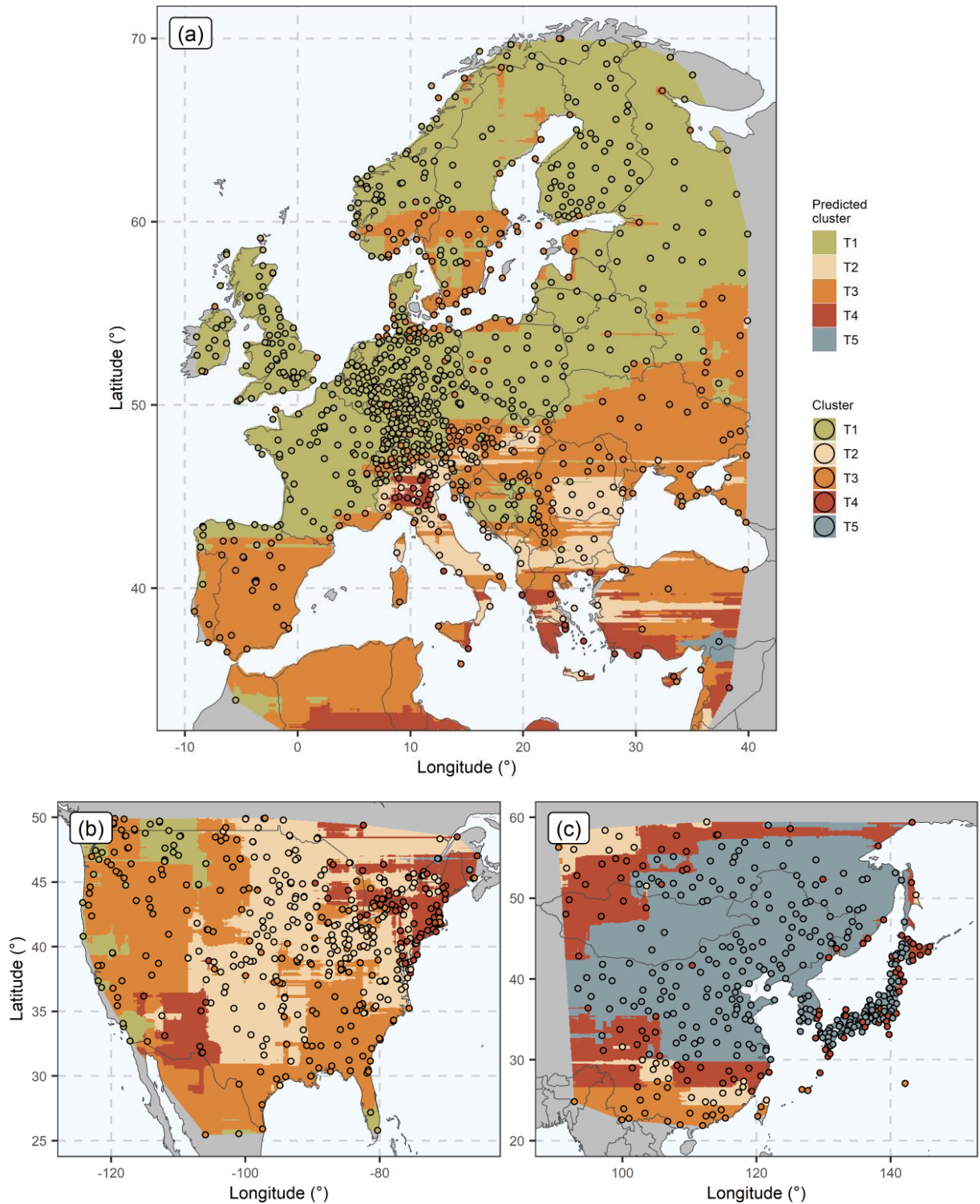

Figure 14. Clusters of the mean monthly temperature and categorical spatial interpolations for (a) Europe and its neighbouring regions, (b) North America, and (c) East Asia.

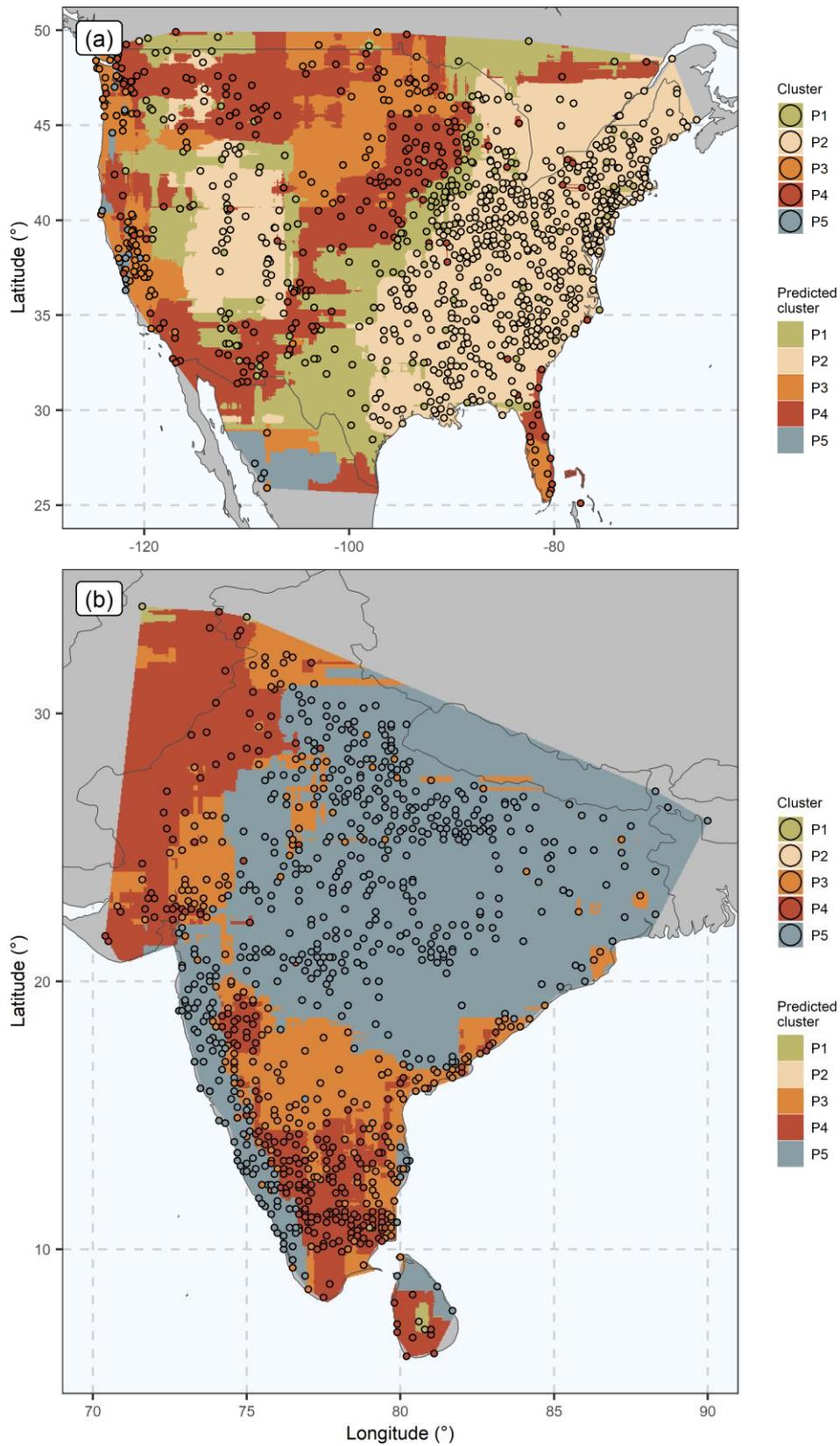

Figure 15. Clustering outcome for the total monthly precipitation time series and outcome of the categorical spatial interpolation for (a) North America and (b) the Indian subcontinent.

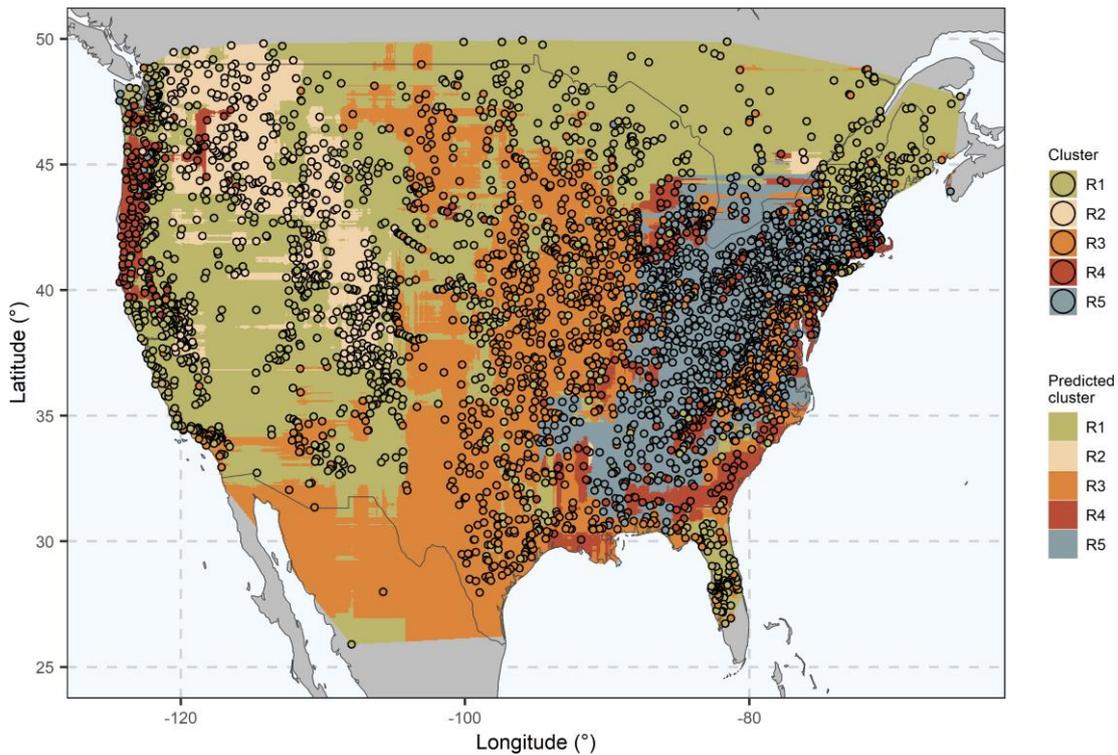

**Figure 16.** Clustering outcome for the mean monthly river flow time series and outcome of the categorical spatial interpolation for North America.

An important remark to be made, at this point, regards the selected number of clusters, which is five. Different numbers of clusters (specified by the user) within the new feature-based time series clustering methodology could lead to somewhat different emerging spatial patterns. Therefore, a more complete investigation of such patterns could involve the repetition of the clustering procedures for various numbers of clusters, and possibly the assessment of the "within-cluster homogeneity" and "between-cluster heterogeneity" by following procedures similar to those proposed by Markonis and Strnad (2020). Although such investigations are in general out of the scope of this work (mostly for reasons of brevity), they can offer interesting insights, and are thus highly recommended. Alternatively, procedures similar to those adopted by Hall and Blöschl (2018) could be incorporated into the new feature-based clustering methodology for automating the selection of an optimal number of clusters according to a predefined criterion. Moreover, additional features (like, e.g., the autocorrelation function derived from the ranks of the original data or the inverse autocorrelation functions suggested by Hamed, 2008, and Hipel and McLeod, 1994, respectively) could possibly enrich the proposed methodological framework, provided that these features complement the aspects already highlighted by the 59 features studied herein. Although this possibility has not been investigated in this work (again for reasons of brevity), it is possible that a somewhat different solution to the three clustering problems may be achieved by using a more extensive feature representation.

Before moving to characterizing the clusters delivered herein in terms of their feature values, it is lastly relevant to explore which are the most important hydroclimatic time series features in solving the three clustering problems, as well as to compare these features with the ones contributing the most to the first principal components obtained for the same datasets (see Figure 4a). For these explorations and comparisons, we present Figure 17. Among the most important features for all three (or at least for two) of the examined main hydroclimatic variable types, as identified by random forests, are the following ones: `x_acf1`, `x_acf10`, `diff1_acf1`, `seas_acf1`, `x_pacf5`, `diff1x_pacf5`, `diff2x_pacf5`, `std1st_der`, `entropy`, `spike` and `seasonal_strength`. Indeed, most of these features are also important from a variance explanation perspective. In general, it seems that the rankings presented in Figure 17 are mostly comparable (though not identical) with those corresponding to the contributions to the first principal components.

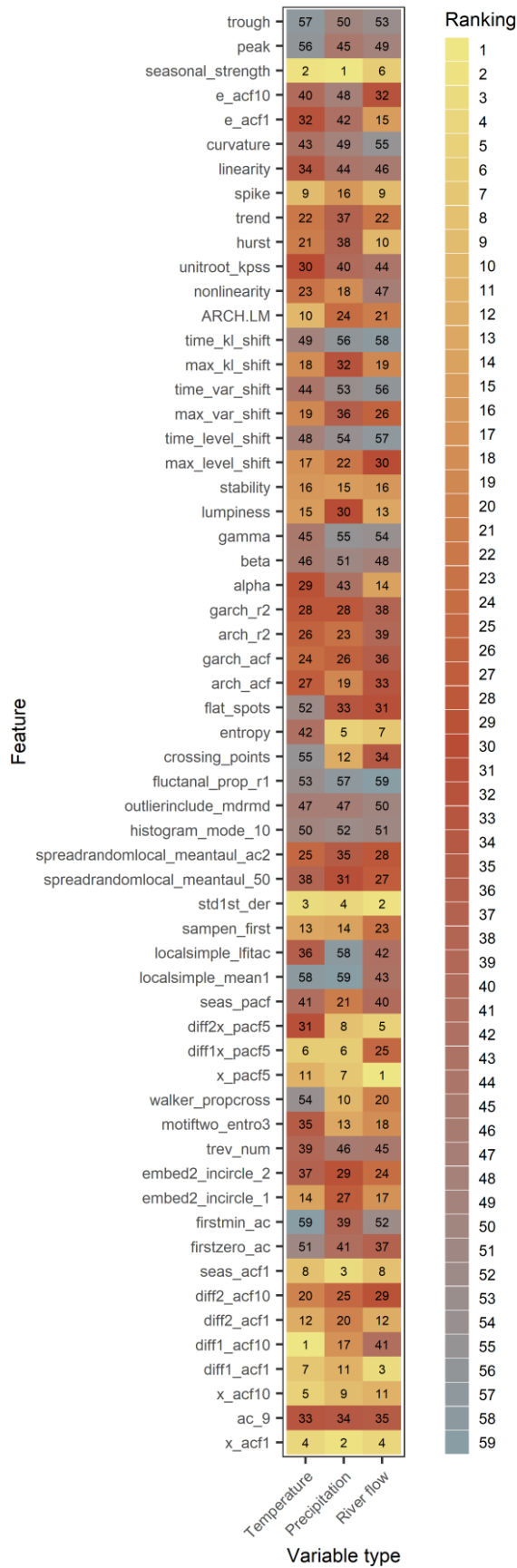

Figure 17. Rankings of the features according to their importance for time series clustering using random forests. The smaller a feature's ranking, the larger its importance. The features are defined in Table 1.

### 3.2.3 Monthly hydroclimatic time series cluster characterizations and analyses

To characterize the hydroclimatic time series clusters obtained by applying the new feature-based methodology, in Figure 18 we present –conditional on the cluster– the violin plots of the top-12 important mean monthly river flow features in time series clustering, as identified by random forests (see Figure 17). Analogous visualizations are also provided for the top-12 important mean monthly temperature features (see Figure S23) and the top-12 important total monthly precipitation features (see Figure S24). In general, we observe that there are overlaps in the ranges of the computed features across the different clusters, with these overlaps being larger or smaller depending on the feature.

An interesting feature for someone to examine in detail is `seasonal_strength`. This feature is, in fact, one of the most interpretable ones (see also the discussions in Section 3.1.1) and its magnitude can be (largely) evident through an optical inspection of the time series. For this inspection, we also present the side-by-side boxplots of the mean monthly river flow values for the twelve months of the year (see Figure 19), while in the Supplement we enclose similar visualizations for mean monthly temperature and total monthly precipitation (see Figures S25 and S26 therein). We first observe in Figure 18f that `seasonal_strength` is mostly larger for cluster R2 and smaller for R3 than for the remaining clusters. We also observe that clusters R4 and R5 are characterized by `seasonal_strength` values with very similar (almost identical) ranges (although they are quite different with respect to other features; see, e.g., Figure 18a–e), and that cluster R1 is characterized by `seasonal_strength` values that cover the entire range of possible values. This information is easily cross-checked by using Figure 19. Specifically, in this latter figure we observe that clusters R4 and R5 are characterized by very similar seasonality patterns, which are also quite intense (see Figure 19d,e), but less intense than seasonality in cluster R2 (see Figure 19b).

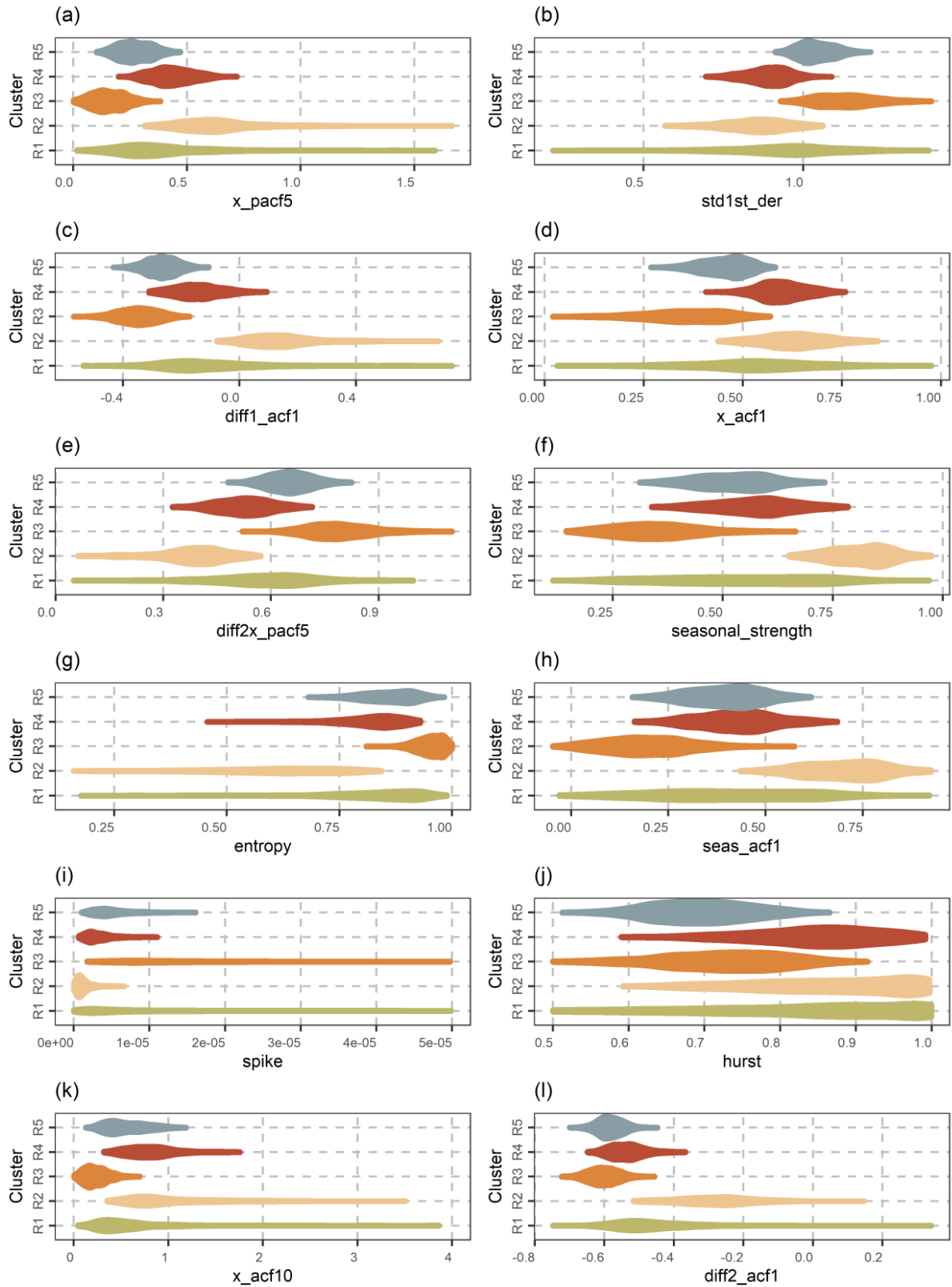

Figure 18. Violin plots of the top-12 important monthly river flow features in time series clustering using random forests conditional on the cluster. The features are presented from (a) the most important to (l) the least important. The corresponding rankings of the features are presented in Figure 17. Outliers have been removed for `spike` and `x_acf10`. The features are defined in Table 1.

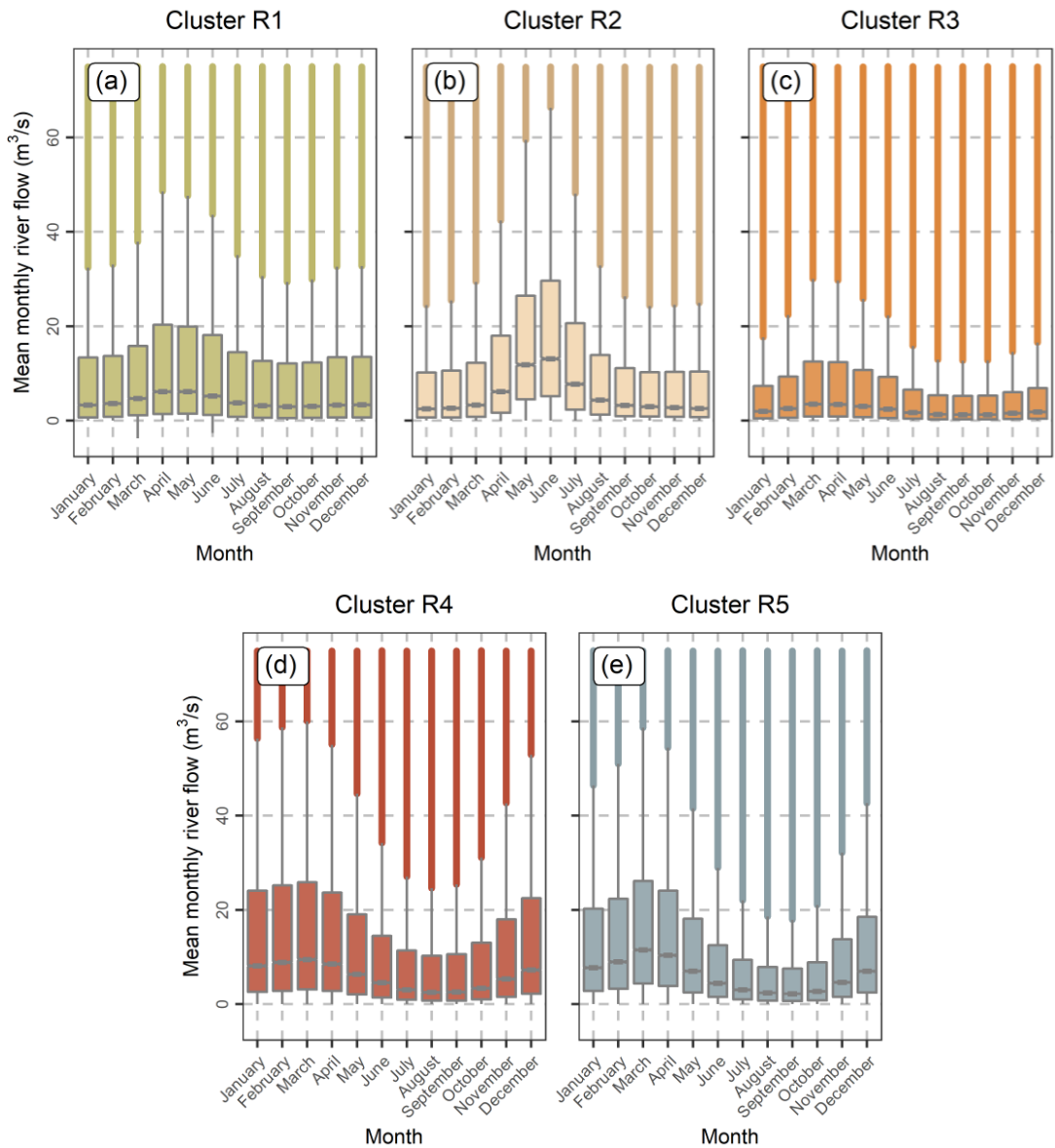

Figure 19. Side-by-side boxplots of the mean monthly river flow values conditional on the cluster and the month. The vertical axes have been truncated at 75 m³/s.

As the data availability time periods are not the same for all stations (since we needed the largest number of 40-year-long time series possible for our experiments), we should lastly check whether the formation of clusters depends on the time period of our time series. For this test, in Figure 20 we present the densities of the last year of the mean monthly river flow time series characterizing the five clusters obtained for mean monthly river flow. These densities show that the formation of clusters is independent of the time period of the time series. Mean monthly temperature and total monthly precipitation clustering exhibit a similar independence (see Figures S27 and S28).

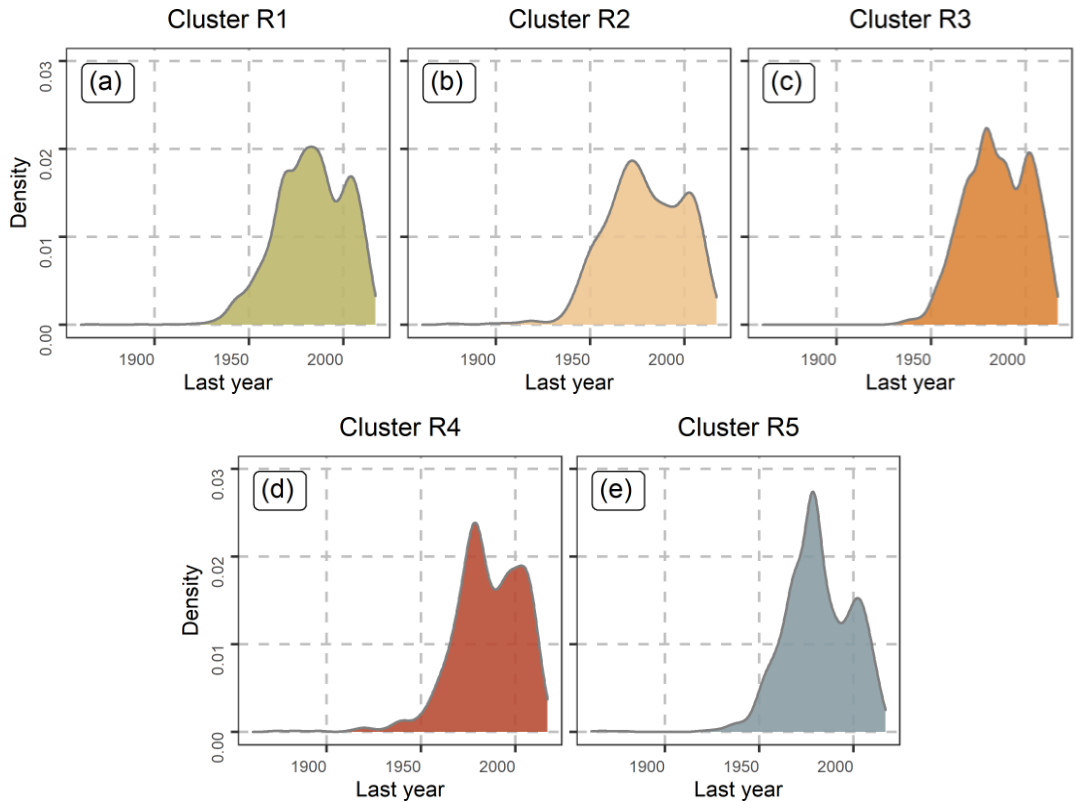

Figure 20. Density plots of the last year of the mean monthly river flow time series conditional on the cluster.

## 4. Further discussion, selected findings and key recommendations

In this section, we further discuss selected findings of the study by emphasizing the implications of the proposed methodological framework (including the new hydroclimatic time series clustering methodology), and by providing additional interpretations and recommendations in line with our research objectives.

### 4.1 On seasonality, trends, temporal dependence, entropy and more

Our premise that a multi-faced and massive approach to time series representation via feature extraction would be meaningful in hydroclimatic contexts has been empirically confirmed by the three global-scale applications of the introduced framework, specifically by the competitive contributions of (most of) the computed features in explaining the total variance of our three big feature datasets (see Section 3.1.2). Moreover, it has led to the delivery of new descriptive insights into the nature of the examined hydroclimatic variables. These insights are particularly important under the concepts of hydrological variability, hydrological change and hydrological similarity, and include those related to the quantification of the similarities and differences observed between the three examined hydroclimatic variable types (see Section 3.1), as well as those related to the identification of homogenous spatial patters characterizing feature variability in space (see Section 3.2). For facilitating these specific quantifications and spatial investigations, we have exploited the entire amount of information resulted from massive feature extraction. This information is summarized in Section 3.1.1 and in the Supplement, and can be used by the interested reader to better understand the computed features and the nature of the three examined hydroclimatic variable types (also by comparing them).

We have also ranked the features according to their contributions in the various principal components (obtained through principal component analyses) and provided ex post knowledge of which of them have been the most informative in solving our three time series clustering problems. Features that have been both (a) identified among the top-10 contributing ones to the first principal components formed for (at least two of) the three feature datasets (see the results presented in Section 3.1.2), and (b) ranked in the first 10 places by the clustering algorithm for (at least two of) the three examined clustering problems (see the results presented in Section 3.2) are the following: `x_acf1`, `x_acf10`, `diff1_acf1`, `seas_acf1`, `x_pacf5`, `std1st_der`, `entropy` and `seasonal_strength`. Most of these features are also quite interpretable. Although the adoption of the entire feature compilation is highly recommended by this study (for its multi-faced and massive character, which resulted in complete hydroclimatic time series characterizations herein), the above-provided ex post knowledge-information may have some practical value for future investigations focusing on monthly temperature, monthly precipitation and monthly river flow. This practical value holds especially for cases in which we cannot afford massive feature extraction solutions, but we are still interested in capturing as much information as possible (given this scale limitation).

By considering this same above-provided information along with the main interests spotted in the hydroclimatic literature (see Section 1.3), in our discussions we have also placed some emphasis on selected interpretable characterizations of seasonality, trends, temporal dependence, entropy, etc. (for reasons of brevity). A feature extensively discussed in this respect is spectral entropy (`entropy`). This feature has been found to vary significantly from region to region for both total monthly precipitation and mean monthly river flow, suggesting regions characterized by lower "forecastability" than others (according to Goerg 2013). Such regions are the greatest part of the examined Australia (identified as the one with the less "forecastable" total monthly precipitation in our created spatial visualizations), a region in the Indian subcontinent (again for total monthly precipitation), approximately half of North America (for its mean monthly river flow) and the largest part of Europe (again for its mean monthly river flow). As it is shown in Papacharalampous and Tyralis (2020) by using approximately 600 mean annual river flow time series and by computing Goerg's (2003) spectral entropy, in situations characterized by larger "forecastability", persistent schemes (e.g., the naïve forecasting scheme simply setting the forecasts equal to the last available observation for the case of non-seasonal processes) are more likely to perform better (compared to sophisticated methodologies) than they are in situations characterized by smaller "forecastability".

Another feature found to be particularly relevant to characterizing hydroclimatic variables in interpretable terms (and in line with the main interests identified in the hydroclimatic literature) is trend strength (`trend`), a practical appealing dimensionless feature that (to our knowledge) has not been exploited before for the investigation of monthly hydroclimatic variables. Although the investigation of trends is quite common in the hydrological and geoscientific literature (Serinaldi et al., 2020a), with many useful descriptive features being exploited in this respect, we believe that this new feature could offer additional descriptive insights and could, therefore, be computed in future works (alongside with other features facilitating the study of trends; see, e.g., Papalexiou and Montanari, 2019; Bertola et al., 2020).

Herein, we have found trend strength to be more intense for mean monthly river flow, while trends have been found to be rather weak (but not negligible) for mean monthly temperature and total monthly precipitation (when contrasted to the magnitude of the "random" variations in the time series). Perhaps, this latter finding is also related to the fact that BATS and Prophet (forecasting methods with trend components) have been found to perform equally well with other sophisticated methods (with no trend components) in the global-scale forecasting investigations focusing on mean monthly temperature and total monthly precipitation by Papacharalampous et al. (2018). Moreover, we have found that the `trend` values of mean monthly river flow that are (roughly) larger than 0.5 are rather scattered across the world (mostly across North America and Europe, since river flow station concentration is larger for these two continental-scale regions). Such spatial patterns are not observed for the remaining features on which we have focused in Section 3.2.1. For instance, quite homogenous spatial patterns characterize the very characteristic feature of sample autocorrelation at lag 1 (`x_acf1`), as well as the spectral entropy (`entropy`) and the seasonality strength (`seasonal_strength`) features.

To our knowledge, investigations focusing on seasonality strength are also new in hydrology and geoscience. We believe that such investigations could offer some additional interpretable insights into the nature of seasonal hydroclimatic variables, along with previously proposed approaches to analysing seasonality (see, e.g., Villarini, 2016; Hall and Blöschl, 2018). Such approaches extensively analyse both the timing and magnitude of seasonal patterns within the year (herein assessed with the side-by-side boxplots of Figure 19), while `seasonal_strength` summarizes in a compact way (i.e., into a single value) and in relative terms information about the magnitude of seasonal fluctuations compared to the magnitude of the "random" fluctuations in the time series (i.e., the component remaining after the removal of the trend and seasonal components).

## 4.2 On the spatial patterns revealed through hydroclimatic time series clustering

Distinct spatially coherent patterns have been identified based on our cluster characterizations (see Section 3.2.2). For instance, most of the mean monthly temperature time series observed in European regions with latitude larger than 45° (or 50°) are attributed to the same cluster. Interestingly, this specific cluster has been shown to mostly characterize these European regions at the global scale. Another continental-scale region identified as interesting to discuss here is East Asia, with the mean monthly temperature time series observed there being mostly attributed to two clusters with spatial homogeneity. Especially, the cluster characterization holding for the largest part of East Asia seems to be extremely rare in other regions across the globe.

Spatial coherence also applies to the cluster characterizations delivered for total monthly precipitation and mean monthly river flow, with the latter hydroclimatic variable type being perhaps the most interesting to examine (because of the large concentration of river flow stations with long observation periods in North America). Notably, total monthly precipitation time series observed in the West Coast of India, a region with tropical monsoon climate, are attributed to a different cluster from those observed in the inner and eastern Indian subcontinent for the same latitudes, while mean monthly river flow time series originating from the western parts of North America and its parts with latitudes from approximately 45° to approximately 50° mostly belong to the same cluster. This cluster is different from the clusters characterizing the mean monthly river flow time series originating from the middle and eastern parts of North America.

The above-outlined information highlights the efficiency of our methods in delivering descriptive and exploratory insights in a completely automatic way (in the sense that the procedures do not depend on the hydroclimatic process at hand). Such insights are important in terms of progressing our understanding of hydroclimatic variable behaviours (primary consideration within our global-scale analyses). Moreover, they build some confidence in using the new time series clustering methodology (and its possible extensions) in the future. Given the spatial coherence characterizing the delivered clusters and the scale-independent nature of the delivered feature values (exploited as inputs to the random forests algorithm), we believe that this methodology could be particularly relevant within regionalization frameworks, in which methodological advances and knowledge of hydrological similarity are exploited in technical and operative terms.

## 5. Summary and conclusions

We have developed a detailed framework to facilitate complete hydroclimatic variable characterizations. This new framework relies on massive feature extraction and four statistical learning (else referred to as "machine learning") algorithms. The adopted feature compilation (composed of 59 diverse features) is supported by past experience and experts' knowledge, which is mostly sourced from scientific fields beyond geoscience and environmental science (e.g., the fields of neuroscience, biology, biomedicine and forecasting), thereby constituting a new concept for our fields. We have empirically proven the high relevance of this new concept in hydrological and hydroclimatic contexts by applying our framework to three global hydroclimatic datasets. These datasets contain 40-year-long information on mean monthly temperature, total monthly precipitation and mean monthly river flow, which originate from over 13 000 stations in total.

Our big data analyses have provided a useful basis for extracting interpretable knowledge (e.g., on seasonality, trends, autocorrelation, long-range dependence, entropy and more, as well as on the relationships between all the computed features) at the global scale, for comparing the examined hydroclimatic variable types in terms of this knowledge, and for identifying distinct patterns related to the features' spatial variability. For this latter purpose, we have also proposed a fully automatic feature-based methodology for hydroclimatic time series clustering, with which we have aspired to exploit a larger part of the information encompassed in the hydroclimatic observations than we could achieve with existing hydroclimatic time series clustering methodologies. Automation is needed in modelling (Chatfield, 1988; Hyndman and Khandakar, 2008; Taylor and Letham, 2018), especially when we are interested in studying problems "at scale" and at a common base. The term "at scale" is used in the literature to imply a large number and variety of problems (see Taylor and Letham, 2018), and is therefore relevant to the study of hydroclimatic variables (since these variables may differ from each other in many ways; see Papalexiou, 2018). In addition to the benefits of our new time series clustering methodology stemming from its automatic nature, the spatially coherent patterns delivered with its use can build some further confidence in its future exploitation for the delivery of descriptive and exploratory insights into the nature of other geophysical variables. Although only such insights have been delivered in the present work, we deem that the technical exploitation of the new methodology within regionalization frameworks could also be beneficial (given the scale-independent nature of the delivered feature values).

We conclude by emphasizing, once again, the massive and automatic character of our methodological framework. With this framework, we believe to have moved a step further from the traditional approach to feature extraction in hydroclimatic research, aiming both to (a) facilitate a better understanding of hydroclimatic variable behaviours, and to (b) deliver more reliable results in hydroclimatic time series clustering contexts.

**Acknowledgements:** We sincerely thank the Associate Editor and the anonymous referees for their very constructive and fruitful suggestions, which helped us to substantially improve this work. Funding from the Italian Ministry of Environment, Land and Sea Protection (MATTM) for SimPRO project (2020–2021) is acknowledged by (in alphabetical order): S. Grimaldi, A. Langousis, G. Papacharalampous, S. M. Papalexiou and E. Volpi. Funding from the Swedish Research Council for Environment, Agricultural Sciences and Spatial Planning (Formas), grant number 2017-00608 is acknowledged by S. Khatami.

## Appendix A   Statistical software information

The analyses and visualizations have been performed in `R` Programming Language (R Core Team, 2020). The following contributed `R` packages have been used: `cluster` (Maechler et al., 2019), `devtools` (Wickham et al., 2020c), `dplyr` (Wickham et al., 2020b), `factoextra` (Kassambara and Mundt, 2020), `forecast` (Hyndman and Khandakar, 2008; Hyndman et al., 2020a), `fracdiff` (Maechler, 2020), `gdata` (Warnes et al., 2017), `geoR` (Ribeiro et al., 2020), `ggbiplot` (Vu, 2011), `ggcorrplot` (Kassambara, 2019), `ggnewscale` (Campitelli, 2020), `ggplot2` (Wickham, 2016a; Wickham et al., 2020a), `gstat` (Gräler et al., 2016; Pebesma and Gräler, 2020), `knitr` (Xie, 2014, 2015, 2020), `lubridate` (Grolemund and Wickham, 2011; Spinu et al., 2020), `maptools` (Bivand and Lewin-Koh, 2020), `maps` (Brownrigg et al., 2018), `randomForest` (Liaw and Wiener, 2002; Liaw, 2018), `readr` (Wickham et al., 2018), `rgdal` (Bivand et al., 2020), `rgeos` (Bivand and Rundel, 2020), `rmarkdown` (Xie et al., 2018; Allaire et al., 2020), `sp` (Pebesma and Bivand, 2005; Bivand et al., 2013; Pebesma and Bivand, 2020), `tidyr` (Wickham, 2020), `tsfeatures` (Hyndman et al., 2020b) and `wesanderson` (Ram and Wickham, 2018).

**Appendix B    Original data sources**

Basic information on real-world data retrieval is provided in Table B1.

Table B1. Basic information on real-world data retrieval. The temperature, precipitation and river flow datasets are documented in Menne et al. (2018), Peterson and Vose (1997), and Do et al. (2018), respectively.

| Data type | Original data source link | Retrieved |
| --- | --- | --- |
| Temperature | https://www.ncdc.noaa.gov/ghcnm/v4.php | 2019-12-29 |
| Precipitation | https://www.ncdc.noaa.gov/ghcnm/v2.php | 2019-12-28 |
| River flow | https://doi.org/10.1594/PANGAEA.887477 | 2019-05-20 |

**Appendix C    Supplement**

The Supplement to this article contains Table S1 and Figures S1–S28. It can be found online at https://doi.org/10.1016/j.scitotenv.2020.144612.

**Declarations of interest:** We declare no conflict of interest.

Montanari, A., Young, G., Savenije, H.H.G., Hughes, D., Wagener, T., Ren, L.L., Koutsoyiannis, D., Cudennec, C., Toth, E., Grimaldi, S., et al., 2013. "Panta Rhei—Everything Flows": Change in hydrology and society—The IAHS Scientific Decade 2013–2022. Hydrological Sciences Journal 58(6), 1256–1275. https://doi.org/10.1080/02626667.2013.809088.

Mukhopadhyay, S., Wang, K., 2020. Breiman's "two cultures" revisited and reconciled. https://arxiv.org/abs/2005.13596.

Nerantzaki, S.D., Papalexiou, S.M., 2019. Tails of extremes: Advancing a graphical method and harnessing big data to assess precipitation extremes. Advances in Water Resources 134, 103448. https://doi.org/10.1016/j.advwatres.2019.103448.

O'Connell, P.E., Koutsoyiannis, D., Lins, H.F., Markonis, Y., Montanari, A., Cohn, T., 2016. The scientific legacy of Harold Edwin Hurst (1880–1978). Hydrological Sciences Journal 61(9), 1571–1590. https://doi.org/10.1080/02626667.2015.1125998.

Pallard, B., Castellarin, A., Montanari, A., 2009. A look at the links between drainage density and flood statistics. Hydrology and Earth System Sciences 13, 1019–1029. https://doi.org/10.5194/hess-13-1019-2009.

Papacharalampous, G., Tyralis, H., 2020. Hydrological time series forecasting using simple combinations: Big data testing and investigations on one-year ahead river flow predictability. Journal of Hydrology 590, 125205. https://doi.org/10.1016/j.jhydrol.2020.125205.

Papacharalampous, G.A., Tyralis, H., Koutsoyiannis, D., 2018. Predictability of monthly temperature and precipitation using automatic time series forecasting methods. Acta Geophysica 66(4), 807–831. https://doi.org/10.1007/s11600-018-0120-7.

Papacharalampous, G.A., Tyralis, H., Koutsoyiannis, D., 2019a. Comparison of stochastic and machine learning methods for multi-step ahead forecasting of hydrological processes. Stochastic Environmental Research and Risk Assessment 33(2), 481–514. https://doi.org/10.1007/s00477-018-1638-6.

Papacharalampous, G.A., Tyralis, H., Langousis, A., Jayawardena, A.W., Sivakumar, B., Mamassis, N., Montanari, A., Koutsoyiannis, D., 2019b. Probabilistic hydrological post-processing at scale: Why and how to apply machine-learning quantile regression algorithms. Water 11(10), 2126. https://doi.org/10.3390/w11102126.

Papalexiou, S.M., 2018. Unified theory for stochastic modelling of hydroclimatic processes: Preserving marginal distributions, correlation structures, and intermittency. Advances in Water Resources 115, 234–252. https://doi.org/10.1016/j.advwatres.2018.02.013.

Papalexiou, S.M., Koutsoyiannis, D., 2012. Entropy based derivation of probability distributions: A case study to daily rainfall. Advances in Water Resources 45, 51–57. https://doi.org/10.1016/j.advwatres.2011.11.007.

Papalexiou, S.M., Koutsoyiannis, D., 2013. Battle of extreme value distributions: A global survey on extreme daily rainfall. Water Resources Research 49(1), 187–201. https://doi.org/10.1029/2012WR012557.

Papalexiou, S.M., Montanari, A., 2019. Global and regional increase of precipitation extremes under global warming. Water Resources Research 55(6), 4901–4914. https://doi.org/10.1029/2018WR024067.

Papalexiou, S.M., AghaKouchak, A., Foufoula-Georgiou, E., 2018a. A diagnostic framework for understanding climatology of tails of hourly precipitation extremes in the United States. Water Resources Research 54(9), 6725–6738. https://doi.org/10.1029/2018WR022732.
62